\documentclass[12pt,preprint]{aastex}

\usepackage{graphicx}
\def\etal{{et al.}}
\def\meszaros{M\'{e}sz\'{a}ros}

\begin{document}

   \title{Analytical light curves in the realistic model for GRB afterglows}

\author{X. F. Wu$^{1}$, Z. G. Dai$^{1}$, Y. F. Huang$^{1}$, T. Lu$^{2}$}
\affil{$^1$Department of Astronomy, Nanjing University, Nanjing
210093, China; Email: xfwu@nju.edu.cn, dzg@nju.edu.cn,
hyf@nju.edu.cn,\\
$^2$Purple Mountain Observatory, Chinese Academy of Sciences,
Nanjing 210008, China; Email: tanlu@mail.pmo.ac.cn}

\begin{abstract}

Afterglow light curves are constructed analytically for realistic
gamma-ray burst remnants decelerating in either a homogeneous
interstellar medium or a stellar wind environment, taking into
account the radiative loss of the blast wave, which affects the
temporal behaviors significantly. Inverse Compton scattering,
which plays an important role when the energy equipartition factor
$\epsilon_{e}$ of electrons  is much larger than that of the
magnetic field ($\epsilon_{B}$), is considered. The inverse
Compton effect prolongs the fast-cooling phase markedly, during
which the relativistic shock is semi-radiative and the radiation
efficiency is approximately constant, $\epsilon=\epsilon_{e}$. It
is further shown that the shock is still semi-radiative for quite
a long time after it transits into the slow-cooling phase, because
of a slow decreasing rate of the radiation efficiency of
electrons. The temporal decaying index of the X-ray afterglow
light curve in this semi-radiative phase is
$(3p-2+2\epsilon)/(4-\epsilon)$ in the interstellar medium case,
and $[3p-2-(p-2)\epsilon]/2(2-\epsilon)$ in the stellar wind case,
where $p$ is the distribution index of the shock-accelerated
electrons. Taking $p=2.2$ --- 2.3 as implied from common shock
acceleration mechanism, and assuming $\epsilon_{e} \sim 1/3$, the
temporal index is more consistent with the observed
$\langle\alpha_{X}\rangle \sim 1.3$ than the commonly used
adiabatic one. The observability of the inverse Compton component
in soft X-ray afterglows is also investigated. To manifest as a
bump or even dominant in the X-ray afterglows during the
relativistic stage, it is required that the density should be
larger than $\sim 1-10$ cm$^{-3}$ in the interstellar medium case,
or the wind parameter $A_{\ast}$ should be larger than $\sim 1$ in
the stellar wind case.

\end{abstract}
\keywords{gamma rays: bursts --- hydrodynamics --- stars: mass loss --- ISM: jets and outflows --- radiation mechanisms: nonthermal --- relativity} 

\section{Introduction}
\label{sec:introduction} In the past seven years, tremendous
progress in understanding the cosmic gamma-ray bursts (GRBs) has
been achieved from the detections of long-lived GRB embers (or
afterglows) in low frequency bands (for reviews see van Paradijs,
Kouveliotou, \& Wijers 2000; Cheng \& Lu 2001; \meszaros\ 2002;
Zhang \& \meszaros\ 2004; Piran 2004). The simple standard shock
model (\meszaros\ \& Rees 1997; Sari, Piran, \& Narayan 1998)
responsible for the afterglow has been successfully established
(Wijers, \meszaros, \& Rees 1997; Vietri 1997; Waxman 1997; Galama
et al. 1998).

As more and more afterglows being detected, the energetics of GRB
remnants and shock microphysics have been inferred within the
context of the standard model. Although the standard model are in
rough agreement with afterglow observations, problems still exist
since the standard scenario is oversimplified at least in two
aspects. First, in the standard afterglow model the relativistic
shock is usually assumed to be quasi-adiabatic. However, the shock
in fact may be partially radiative. Fittings to observed
afterglows reveal that the shock imparts an equipartition amount
of energy into electrons, which is responsible for both the
afterglow emission and the shock energy loss (Panaitescu \& Kumar
2001, 2002). The temporal evolution of the shock energy would
affect the estimation of the GRB energetics from late time
afterglows, as well as the profiles of afterglow light curves
(Lloyd-Ronning \& Zhang 2004; B\"{o}ttcher \& Dermer 2000).
Second, multi-wavelength fittings to several afterglows also
indicate that post-shock energy density imparted to electrons is
statistically much larger than that of the magnetic fields
(Panaitescu \& Kumar 2001). It implies that the inverse Compton
(IC) scattering plays an important role in GRB afterglows. The IC
scattering has two effects. One is to enhance the cooling rate of
the shock-accelerated electrons and hence delay the transition
from early fast-cooling phase to late slow-cooling phase. It
finally influences the observed flux density. Another effect is to
cause a high energy spectral component typically above the soft
X-ray band. These IC effects have been taken into account by
numerous authors (Waxman 1997; Wei \& Lu 1998, 2000; Panaitescu \&
Kumar 2000; Dermer, B\"{o}ttcher, \& Chiang 2000; Sari \& Esin
2001; Bj\"{o}rnsson 2001; Wang, Dai, \& Lu 2001; Zhang \&
\meszaros\ 2001; Li, Dai, \& Lu 2002).

As for the circum-burst environment, in the standard model it is
once assumed that the surroundings of GRBs are homogeneous
interstellar medium (ISM). Over the past several years, a lot of
evidence has been collected linking GRBs to the core collapse of
massive stars (Woosley 1993; Paczy\'{n}ski 1998). The most
important evidence came from the direct association between GRB
$030329$ and the supernova SN $2003$dh (Stanek et al. 2003; Hjorth
et al. 2003), as well as the previous tentative association
between GRB $980425$ and SN $1998$bw (Kulkarni et al. 1998; Galama
et al. 1998). These associated supernovae have been confirmed to
be type Ib/c SNe. The progenitors of type Ib/c SNe are commonly
recognized as massive Wolf-Rayet stars. During their whole life,
the massive progenitors eject their envelope material into their
surroundings through line pressure and thus the stellar wind
environments are formed. This means that the circum-burst medium
for GRB afterglows may be the stellar wind (Dai \& Lu 1998;
\meszaros, Rees, \& Wijers 1998; Chevalier \& Li 1999, 2000;
Panaitescu \& Kumar 2000).

In this paper, we study the afterglow properties of realistic GRB
shocks, considering the effect of energy losses. The circum-burst
environment is assumed to be either the ISM-type or the stellar
wind type. We present an analytical solution for the realistic
blast wave during the fast-cooling phase of GRB afterglows in \S
\ref{sec:fastcooling}. This semi-radiative hydrodynamics is
applied to the late slow-cooling phase with quite reasonable
argument in \S \ref{sec:slowcooling}. Constraints on the IC
components in the soft X-ray afterglows are given in both
sections. In \S \ref{sec:lightcurves} we illustrate typical
analytical light curves for the realistic model in detail.
Conclusions and discussion are presented in \S
\ref{sec:conclusion}.

\section{The early fast-cooling phase}
\label{sec:fastcooling} The realistic model for GRB remnants has
been extensively investigated in the past few years (Huang, Dai, \&
Lu 1999; Huang et al. 2000). It has been shown that this model is
correct for both adiabatic and radiative fireballs, and in
both ultra-relativistic and non-relativistic phases. The basic
hydrodynamic equation of this model can be derived as follows. In
the fixed frame, which is rest to the circum-burst environment,
the total kinetic energy of the fireball is
$E_{\rm{K}}=(\gamma-1)(M_{\rm{ej}}+M_{\rm{sw}})c^{2}+(1-\epsilon)\gamma
U$, where $\gamma$ is the Lorentz factor of the blast wave,
$M_{\rm{ej}}$ is the initial mass of the blast wave ejected from
the central engine, $M_{\rm{sw}}$ is the mass of the swept-up
ambient medium, $c$ is the speed of light, and $\epsilon$ is the
total radiation efficiency (Huang et al. 1999). In the comoving
frame of the blast wave, the total internal energy instantaneously
heated by the shock is $U=(\gamma-1)M_{\rm{sw}}c^2$, which is
implied from the relativistic jump conditions (Blandford \& McKee
1976). The differential loss of the kinetic energy $E_{\rm{K}}$,
when the blast wave sweeps up an infinitesimal mass
$\rm{d}\it{M_{\rm{sw}}}$, can
be formulated as
\begin{equation}
\rm{d}[(\gamma-1)(\textit{M}_{\rm{ej}}+\textit{M}_{\rm{sw}})\textit{c}^{2}+(1-\epsilon)\gamma
\textit{U}]=-\epsilon\gamma(\gamma-1)\rm{d}\textit{M}_{\rm{sw}}\textit{c}^{2}.
\end{equation}
Assuming a constant $\epsilon$ and inserting the expression for
$U$, it is then easy to obtain the hydrodynamic equation of the realistic model
(Huang et al. 1999, 2000)
\begin{equation}
\frac{\rm{d}\gamma}{\rm{d}\it{M_{\rm{sw}}}}=-\frac{\gamma^2-1}{M_{\rm{ej}}+\epsilon
M_{\rm{sw}}+2(1-\epsilon)\gamma M_{\rm{sw}}}. \label{eqn:hydro}
\end{equation}
Feng et al. (2002) have relaxed the assumption of a constant
$\epsilon$, and found that the results
differ little from the above equation. As we show
later, the epoch for a constant radiation efficiency will not end
just at the transition from the fast cooling phase to the slow cooling
phase of the fireball evolution. In fact, it would last for a time
much longer than that transition time, because the radiation
efficiency of the shock-accelerated electrons in the slow cooling
phase decreases very slowly for typical values of
the index of electron energy distribution, e.g. $p\approx2.2$
indicated both from observations of the afterglow spectra, and
from the shock acceleration theory (Achterberg et al. 2001).

Throughout this paper, we focus on the early epoch when the
afterglow of a relativistic jet is spherical-like, which requires
the Lorentz factor of the jet ($\gamma$) being larger than the
inverse of the half-opening angle. The swept-up mass is given by
$M_{\rm{sw}}=\displaystyle\frac{4\pi}{3-k}m_{p}nR^{3}$, where
$m_{p}$ is the proton mass, and the ambient density is
\begin{equation}
n=AR^{-k}, \label{eqn:density}
\end{equation}
where $k=0$ with $n=A=\rm{const}$ for the ISM case, and $k=2$ with
$A=3\times 10^{35}A_{\ast}$ cm$^{-1}$ for the stellar wind case.
Such a blast wave begins to decelerate at the radius
$R_{\rm{dec}}$, when the swept-up mass reaches
$M_{\rm{ej}}/\gamma_0$, where $\gamma_0$ is the
initial Lorentz factor. The corresponding decelerating time
measured by an observer is
$t_{\rm{dec}}=R_{\rm{dec}}(1+z)/2\gamma_0^2 c$, here $z$ is
the cosmological redshift of the GRB. We neglect the effect of
reverse shocks in the early afterglow for simplicity.

At early times, electrons cool rapidly.
The blast wave is therefore semi-radiative with a
constant radiation efficiency $\epsilon=\epsilon_{e}$. The typical
energy equipartition factor of electrons is $\epsilon_{e}\sim
1/3$. Equation (\ref{eqn:hydro}) can then be analytically integrated by
neglecting the first two terms in the denominator at the right
side  when $t>t_{\rm{dec}}$. The
scaling laws for the hydrodynamics are $\gamma^2\propto R^{-m}$
and $R\propto t^{1/(m+1)}$, where the hydrodynamic self-similarity
index
\begin{equation}
m=\displaystyle\frac{3-k}{1-\epsilon}, \label{eqn:m}
\end{equation}
and $t$ is the observed time since the burst. The $\epsilon$ term in
the denominator in equation (\ref{eqn:m}) shows the deviation of
the hydrodynamics of a semi-radiative blast wave from that of an
adiabatic one of Blandford \& McKee (1976). Since the
isotropic-equivalent energy $E$ is proportional to
$M_{\rm{sw}}\gamma^2$, it decreases as
\begin{equation}
E=E_{\rm{dec}}(\frac{t}{t_{\rm{dec}}})^{-(3-k)\epsilon/(4-k-\epsilon)},
\label{eqn:energy}
\end{equation}
where $E_{\rm{dec}}$ is the initial isotropic energy at
$R_{\rm{dec}}$. The minimum Lorentz factor of the
shock-accelerated electrons is evaluated by
\begin{equation}
\gamma_{e,\rm{min}}=\frac{\epsilon_{e}}{6}\frac{m_{p}}{m_{e}}\zeta_{1/6}\gamma_{0}(\frac{t}{t_{\rm{dec}}})^{-m/[2(m+1)]},
\label{eqn:gm}
\end{equation}
where $\zeta_{1/6}=6\displaystyle\frac{p-2}{p-1}$ and $m_{e}$ is
the electron mass. Conventionally, assuming a constant fraction of
$\epsilon_{B}$ of the post-shock thermal energy density contained
in post-shock magnetic fields, the magnetic field intensity
in the comoving frame is
\begin{equation}
B=B_{\rm{dec}}(\frac{t}{t_{\rm{dec}}})^{-(m+k)/[2(m+1)]},
\label{eqn:B}
\end{equation}
where the initial value is $B_{\rm{dec}}=(32\pi \epsilon_{B}A
R_{\rm{dec}}^{-k}m_{p}c^2)^{1/2}\gamma_{0}$. The maximum Lorentz
factor of the shock-accelerated electrons,
$\gamma_{e,\rm{max}}\approx 10^{8}(B/\rm{G})^{-1/2}$, is obtained
by assuming that the acceleration timescale, which is typically
the gyration period in the magnetic field, equals the
hydrodynamical timescale. The cooling Lorentz factor of electrons
is determined by considering both synchrotron radiation and inverse Compton
scattering, i.e. (Sari et al. 1998; Panaitescu \& Kumar 2000)
\begin{equation}
\gamma_{c}=\frac{6\pi
m_{e}c(1+z)}{\sigma_{\rm{T}}B_{\rm{dec}}^2\gamma_0
t_{\rm{dec}}(1+Y)}(\frac{t}{t_{\rm{dec}}})^{(m+2k-2)/[2(m+1)]},
\label{eqn:fc:gc}
\end{equation}
where $\sigma_{\rm{T}}$ is the Thomson cross section, and the
Compton parameter
$Y=(\displaystyle\frac{\epsilon_{e}}{\epsilon_{B}})^{1/2}\gg 1$ is
a constant during the fast cooling phase (Sari \& Esin 2001). The
transition from the fast-cooling phase to the slow-cooling phase
happens at $t_{cm}$, when $\gamma_{c}=\gamma_{e,\rm{min}}$.
Combining equations (\ref{eqn:m}) --- (\ref{eqn:fc:gc}) with the
definition of the deceleration time, we obtain
\begin{equation}
t_{cm}=\frac{(1+z)\sigma_{\rm{T}}^{1/2}}{2c}A\sigma_{\rm{T}}^{(3-k)/2}[\frac{(3-k)E_{cm}}{4\pi
m_{p}c^2}]^{(2-k)/2}(\frac{2m_{p}}{3m_{e}}\epsilon_{e}^{3/4}\epsilon_{B}^{1/4}\zeta_{1/6}^{1/2})^{4-k}.
\label{eqn:t0-1}
\end{equation}
Here the subscript ``cm'' denotes the physical quantity at the
time when $\gamma_{c}=\gamma_{e,\rm{min}}$. Throughout this paper
we are especially interested in the usual case of $\epsilon_{e}\gg\epsilon_{B}$, in
which the inverse Compton scattering has a dominant effect on the
evaluation of $\gamma_{c}$. We denote the isotropic energy at
$t_{cm}$ as $E_{cm}=E(t_{cm})$. Note that $t_{cm}$ is independent
of the radiation efficiency $\epsilon$ and the initial Lorentz
factor $\gamma_{0}$. The value of $t_{cm}$ can be further
calculated to be
\begin{equation}
t_{cm}=\left\{ \begin{array}{l} 0.30(1+z)\epsilon_{e,-0.5}^3
\epsilon_{B,-2.5}\zeta_{1/6}^{2}E_{cm,53}n\,\,\rm{day},   \phantom{sss}  \rm{ISM,} \\
0.58(1+z)\epsilon_{e,-0.5}^{3/2}
\epsilon_{B,-2.5}^{1/2}\zeta_{1/6}A_{\ast}\,\,\rm{day},  \;\phantom{sssssss}  \rm{wind.}\\
 \end{array} \right.
 \label{eqn:t0-2}
\end{equation}
Here we adopt the conventional definition of $Q=Q_{x}10^{x}$.
The radius of the blast wave at $t_{cm}$ is
\begin{equation}
R_{cm}=\frac{2m_{p}}{3m_{e}}\epsilon_{e}^{3/4}\epsilon_{B}^{1/4}\zeta_{1/6}^{1/2}[\frac{(3-k)E_{cm}}{4\pi
m_{p}c^2}]^{1/2}\sigma_{\rm{T}}^{1/2}, \label{eqn:Radius-1}
\end{equation}
or equivalently
\begin{equation}
R_{cm}=\left\{ \begin{array}{l}
3.98\times10^{17}\epsilon_{e,-0.5}^{3/4}
\epsilon_{B,-2.5}^{1/4}\zeta_{1/6}^{1/2}E_{cm,53}^{1/2}\,\,\rm{cm},   \phantom{sss}  \rm{ISM,} \\
2.30\times10^{17}\epsilon_{e,-0.5}^{3/4}
\epsilon_{B,-2.5}^{1/4}\zeta_{1/6}^{1/2}E_{cm,53}^{1/2}\,\,\rm{cm},   \phantom{sss}  \rm{wind.} \\
 \end{array} \right.
 \label{eqn:Radius-2}
\end{equation}
The evolution of the radius is
$R=R_{cm}\displaystyle(\frac{t}{t_{cm}})^{(1-\epsilon)/(4-k-\epsilon)}$.
The magnetic field at $t_{cm}$ is
\begin{equation}
B_{cm}=3(\frac{3}{2\pi})^{1/4}\frac{m_{e}^{2}c^{4}}{e^{3}}\frac{3m_{e}}{2m_{p}}\epsilon_{e}^{-9/8}\epsilon_{B}^{1/8}\zeta_{1/6}^{-3/4}[\frac{(3-k)E_{cm}}{4\pi
m_{p}c^2}]^{-1/4}, \label{eqn:B-1}
\end{equation}
or numerically
\begin{equation}
B_{cm}=\left\{ \begin{array}{l} 0.35\epsilon_{e,-0.5}^{-9/8}
\epsilon_{B,-2.5}^{1/8}\zeta_{1/6}^{-3/4}E_{cm,53}^{-1/4}\,\,\rm{G},   \phantom{sss}  \rm{ISM,} \\
0.46\epsilon_{e,-0.5}^{-9/8}
\epsilon_{B,-2.5}^{1/8}\zeta_{1/6}^{-3/4}E_{cm,53}^{-1/4}\,\,\rm{G},   \phantom{sss}  \rm{wind.} \\
 \end{array} \right.
 \label{eqn:B-2}
\end{equation}
The magnetic field evolves as
$B=B_{cm}\displaystyle(\frac{t}{t_{cm}})^{-(3-k\epsilon)/[2(4-k-\epsilon)]}$.

\subsection{Properties of the synchrotron radiation}
The characteristic synchrotron frequencies corresponding to the
$\gamma_{c}$, $\gamma_{e,\rm{min}}$ and $\gamma_{e,\rm{max}}$
electrons are denoted as $\nu_{c},\nu_{m}$ and $\nu_{M}$ respectively.
They can be easily calculated according to
$\nu =\displaystyle\gamma\gamma_{e}^2\frac{eB}{2\pi(1+z)m_{e}c}$,
with $e$ being the electron charge. The peak flux
density of the afterglow is
$F_{\nu,\rm{max}}=\displaystyle\frac{N_{\rm{e}}P_{\nu,\rm{max}}}{4\pi
D_{\rm{L}}^{2}}(1+z)$, where
$N_{\rm{e}}=\displaystyle\frac{4\pi}{3-k}AR^{3-k}$ is the total number of
shocked accelerated electrons,
$P_{\nu,\rm{max}}=\displaystyle\frac{\sigma_{\rm{T}}m_{e}c^2}{3e}\gamma
B$ is the peak spectral power of a single electron and
$D_{\rm{L}}$ is the luminosity distance (Sari et al. 1998). This
peak flux density is at the cooling frequency $\nu_{c}$ in
the fast-cooling phase, and the flux density at $\nu_{c}$ would be
reduced if the synchrotron-self-absorption (SSA) frequency
$\nu_{a}$ is above $\nu_{c}$.

It is convenient to re-scale the physical quantities to the values
at the time $t_{cm}$, because physical variables such as
$\gamma_{c}$ and $\gamma_{m}$ at $t_{cm}$ are independent of
$\epsilon$. The minimum electron Lorentz factor equals to the
cooling Lorentz factor at $t_{cm}$,
$\gamma_{e,cm}\equiv\gamma_{e,\rm{min}}(t_{cm})=\gamma_{c}(t_{cm})$,
\begin{equation}
\gamma_{e,cm}=\displaystyle\frac{1}{4}(\frac{2m_{p}}{3m_{e}})^{(k-1)/2}\epsilon_{e}^{(3k-1)/8}\epsilon_{B}^{(k-3)/8}
                     \zeta_{1/6}^{(k+1)/4}(A\sigma_{\rm{T}}^{(3-k)/2})^{-1/2}[\frac{(3-k)E_{cm}}{4\pi m_{p}c^{2}}]^{(k-1)/4},
\label{eqn:gamma_cm-1}
\end{equation}
which can be evaluated to be
\begin{equation}
\gamma_{e,cm}=\left\{ \begin{array}{l}
1536\epsilon_{e,-0.5}^{-1/8}
\epsilon_{B,-2.5}^{-3/8}\zeta_{1/6}^{1/4}E_{cm,53}^{-1/4}n^{-1/2},   \phantom{sss}  \rm{ISM,} \\
849\epsilon_{e,-0.5}^{5/8}
\epsilon_{B,-2.5}^{-1/8}\zeta_{1/6}^{3/4}E_{cm,53}^{1/4}A_{\ast}^{-1/2},  \phantom{ssss}  \rm{wind.}\\
 \end{array} \right.
\label{eqn:gamma_cm-2}
\end{equation}
The typical frequency $\nu_{m}$ also equals to the cooling
frequency $\nu_{c}$ at $t_{cm}$, i.e.
$\nu_{cm}\equiv\nu_{m}(t_{cm})=\nu_{c}(t_{cm})$, which is
\begin{eqnarray}
\nu_{cm}=&&\displaystyle\frac{1}{16}(\frac{3}{2\pi})^{5/4}\frac{m_{e}c^3}{e^2(1+z)}(\frac{2m_{p}}{3m_{e}})^{(3k-7)/2}\epsilon_{e}^{(9k-20)/8}\epsilon_{B}^{(3k-8)/8}
            \zeta_{1/6}^{(3k-4)/4}\nonumber\\
         &&\times(A\sigma_{\rm{T}}^{(3-k)/2})^{-3/2}[\frac{(3-k)E_{cm}}{4\pi
m_{p}c^{2}}]^{(3k-4)/4}, \label{eqn:nu_0-1}
\end{eqnarray}
and can be further deduced to be
\begin{equation}
\nu_{cm}=\left\{ \begin{array}{l}
3.65\times 10^{13}(1+z)^{-1}\epsilon_{e,-0.5}^{-5/2}\epsilon_{B,-2.5}^{-1}\zeta_{1/6}^{-1}E_{cm,53}^{-1}n^{-3/2}\,\,\rm{Hz}, \phantom{sss}  \rm{ISM,} \\
8.10\times 10^{12}(1+z)^{-1}\epsilon_{e,-0.5}^{-1/4}\epsilon_{B,-2.5}^{-1/4}\zeta_{1/6}^{1/2}E_{cm,53}^{1/2}A_{\ast}^{-3/2}\,\,\rm{Hz},  \phantom{ss}  \rm{wind.}\\
 \end{array} \right.
\label{eqn:nu_0-2}
\end{equation}
The maximum frequency of the synchrotron radiation at $t_{cm}$ is
\begin{equation}
\nu_{M}(t_{cm})=\left\{ \begin{array}{l}
4.3\times 10^{25}(1+z)^{-1}\epsilon_{e,-0.5}^{-1/8}\epsilon_{B,-2.5}^{-3/8}\zeta_{1/6}^{1/4}E_{cm,53}^{-1/4}n^{-1/2}\,\,\rm{Hz}, \phantom{ss}  \rm{ISM,} \\
2.4\times 10^{25}(1+z)^{-1}\epsilon_{e,-0.5}^{5/8}\epsilon_{B,-2.5}^{-1/8}\zeta_{1/6}^{3/4}E_{cm,53}^{1/4}A_{\ast}^{-1/2}\,\,\rm{Hz},  \phantom{ss}  \rm{wind,}\\
 \end{array} \right.
\label{eqn:nu_max}
\end{equation}
which corresponds to $\sim 100$ GeV photons. It ensures that the
synchrotron spectrum can be extrapolated to very high energy band,
as will be useful in the next subsection. The peak flux density of
the synchrotron radiation at $t_{cm}$ is
\begin{eqnarray}
F_{\nu,\rm{max}}(t_{cm})=&&\displaystyle(\frac{2\pi}{3})^{3/4}\frac{4m_{e}c^{2}(1+z)}{(3-k)D_{\rm{L}}^{2}}(\frac{3m_{e}}{2m_{p}})^{(k-1)/2}\epsilon_{e}^{-3k/8}\epsilon_{B}^{(4-k)/8}
                     \zeta_{1/6}^{-k/4}\nonumber\\
                 &&\times(A\sigma_{\rm{T}}^{(3-k)/2})^{1/2}[\frac{(3-k)E_{cm}}{4\pi
m_{p}c^{2}}]^{(4-k)/4}, \label{eqn:Fnu_max-1}
\end{eqnarray}
which is numerically expressed as
\begin{equation}
F_{\nu,\rm{max}}(t_{cm})=\left\{ \begin{array}{l}
44(1+z)\epsilon_{B,-2.5}^{1/2}E_{cm,53}n^{1/2}D_{\rm{L},28}^{-2}\,\,\rm{mJy}, \;\phantom{ssssssssssss}  \rm{ISM,} \\
104(1+z)\epsilon_{e,-0.5}^{-3/4}\epsilon_{B,-2.5}^{1/4}\zeta_{1/6}^{-1/2}E_{cm,53}^{1/2}A_{\ast}^{1/2}D_{\rm{L},28}^{-2}\,\,\rm{mJy},  \phantom{s}  \rm{wind.}\\
 \end{array} \right.
\label{eqn:Fnu_max-2}
\end{equation}
We obtain the temporal evolutions of these characteristic
frequencies and the peak flux density during the fast-cooling
phase as follows,
\begin{eqnarray}
&&\nu_{c}=\nu_{cm}(\frac{t}{t_{cm}})^{(3k-4)/[2(m+1)]},\phantom{sssssssss}
\nu_{m}=\nu_{cm}(\frac{t}{t_{cm}})^{-(4m+k)/[2(m+1)]},\nonumber\\
&&\nu_{M}=\nu_{M}(t_{cm})(\frac{t}{t_{cm}})^{-m/[2(m+1)]},\phantom{ss}
F_{\nu,\rm{max}}=F_{\nu,\rm{max}}(t_{cm})(\frac{t}{t_{cm}})^{(6-3k-2m)/[2(m+1)]}.
\label{eqn:fc:spectrum}
\end{eqnarray}

The synchrotron-self-absorption frequency in the fast cooling
phase can be evaluated by
\begin{equation}
\nu_{a}=\left\{ \begin{array}{l}
\nu_{a,<}\equiv\nu_{c}\displaystyle[\frac{c_{0}}{(3-k)}\frac{enR}{B\gamma_{c}^{5}}]^{3/5}, \phantom{ss} \rm{if} \phantom{ss} \nu_{a}<\nu_{c}, \\
\nu_{a,>}\equiv\nu_{c}\displaystyle[\frac{c_{0}}{(3-k)}\frac{enR}{B\gamma_{c}^{5}}]^{1/3}, \phantom{ss} \rm{if} \phantom{ss} \nu_{c}<\nu_{a}<\nu_{m},\\
 \end{array} \right.
 \label{eqn:fc:nu_a-1}
\end{equation}
where $c_{0}\approx10.4\displaystyle\frac{p+2}{p+2/3}$ (see the
appendix of Wu et al. 2003). When $t<t_{cm}$, the distribution of
the cooled electron with
$\gamma_{c}<\gamma_{\rm{e}}<\gamma_{e,\rm{min}}$ has $p=2$ and the
resulting coefficient $c_{0}=15.6$. For $2<p<3$, the value of
$c_{0}$ is nearly a constant, $\sim 15$, which is about $3$ times
larger than the coefficient in equation (52) of Panaitescu \&
Kumar (2000). The SSA frequency can be determined straightforwardly
by
\begin{equation}
\nu_{a}=\rm{min}\{\it{\nu_{a,<},\nu_{a,>}}\}
 \label{eqn:fc:nu_a-2}
\end{equation}
without judging whether $\nu_{a}<\nu_{c}$ or not. The case for
$\nu_{a}>\nu_{m}$ can be neglected. The numerical expression for
$\nu_{a,<}$ is
\begin{equation}
\nu_{a,<}=\left\{ \begin{array}{l}
4.75(1+z)^{-1}\epsilon_{e,-0.5}^{-1}\epsilon_{B,-2.5}^{1/5}\zeta_{1/6}^{-1}E_{cm,53}^{1/5}n^{3/5}(\displaystyle\frac{t}{t_{cm}})^{-(10+8\epsilon)/[5(4-\epsilon)]}\,\,\rm{GHz}, \,\phantom{sss}  \rm{ISM,} \\
20.9(1+z)^{-1}\epsilon_{e,-0.5}^{-19/10}\epsilon_{B,-2.5}^{-1/10}\zeta_{1/6}^{-8/5}E_{cm,53}^{-2/5}A_{\ast}^{3/5}(\displaystyle\frac{t}{t_{cm}})^{-(16-10\epsilon)/[5(2-\epsilon)]}\,\,\rm{GHz},  \phantom{}  \rm{wind,}\\
 \end{array} \right.
 \label{eqn:fc:nu_a<}
\end{equation}
while the expression for $\nu_{a,>}$ is
\begin{equation}
\nu_{a,>}=\left\{ \begin{array}{l}
2.53\times10^{11}(1+z)^{-1}\epsilon_{e,-0.5}^{-5/3}\epsilon_{B,-2.5}^{-1/3}\zeta_{1/6}^{-1}E_{cm,53}^{-1/3}n^{-1/3}(\displaystyle\frac{t}{t_{cm}})^{-2/(4-\epsilon)}\,\,\rm{Hz}, \phantom{sss}  \rm{ISM,} \\
2.96\times10^{11}(1+z)^{-1}\epsilon_{e,-0.5}^{-7/6}\epsilon_{B,-2.5}^{-1/6}\zeta_{1/6}^{-2/3}A_{\ast}^{-1/3}(\displaystyle\frac{t}{t_{cm}})^{-2/3}\,\,\rm{Hz},  \phantom{ssssssssss}  \rm{wind.}\\
 \end{array} \right.
  \label{eqn:fc:nu_a>}
\end{equation}
For the ISM case, the transition from initially
$\nu_{a}=\nu_{a,>}$ to the later $\nu_{a}=\nu_{a,<}$ happens when
$\nu_{a}=\nu_{c}$,
\begin{equation}
t_{ac}=\left\{ \begin{array}{l}
6.03\times10^{-51}\epsilon_{e,-1}^{16.25}\epsilon_{B,-2.5}^{13}E_{cm,53}^{13}n^{22.75}t_{cm}, \phantom{ss}  \,\rm{if} \phantom{ss} \epsilon=0.1, \\
2.67\times10^{-13}\epsilon_{e,-0.5}^{4.85}\epsilon_{B,-2.5}^{3.88}E_{cm,53}^{3.88}n^{6.8}t_{cm}, \phantom{sss}  \rm{if} \phantom{ss} \epsilon=0.32,\\
\end{array} \right.
\label{eqn:fc:ISM:t_ac}
\end{equation}
which indicates that the epoch when $\nu_{a}=\nu_{a,>}$ is very
short, unless the ISM is very dense, e.g. $n\gtrsim10^2$ (Dai \&
Lu 1999, 2000). For the stellar wind case, the transition from
$\nu_{a}=\nu_{a,>}$ to $\nu_{a}=\nu_{a,<}$ takes place when
$\nu_{a}=\nu_{c}$,
\begin{equation}
t_{ac}=\left\{ \begin{array}{l}
0.14\epsilon_{e,-1}^{-0.80}\epsilon_{B,-2.5}^{0.07}\zeta_{1/6}^{-1.02}E_{cm,53}^{-0.44}A_{\ast}^{1.02}t_{cm}, \phantom{sss}  \rm{if} \phantom{ss} \epsilon=0.1, \\
0.046\epsilon_{e,-0.5}^{-0.85}\epsilon_{B,-2.5}^{0.08}\zeta_{1/6}^{-1.09}E_{cm,53}^{-0.47}A_{\ast}^{1.09}t_{cm}, \;\phantom{s}  \rm{if} \phantom{ss} \epsilon=0.32.\\
\end{array} \right.
\label{eqn:fc:wind:t_ac}
\end{equation}

The flux density at the observed frequency $\nu$ from the
synchrotron component for $t<t_{ac}$ is
\begin{equation}
F_{\nu}=\left\{
\begin{array}{l}
\displaystyle(\frac{\nu}{\nu_{a}})^{2}(\frac{\nu_{c}}{\nu_{a}})F_{\nu,\rm{max}}\propto
t^{(1+k+m)/(m+1)},   \phantom{sssssssssssssssss}  \nu<\nu_{c}, \\
\displaystyle(\frac{\nu}{\nu_{a}})^{5/2}(\frac{\nu_{a}}{\nu_{c}})^{-1/2}F_{\nu,\rm{max}}\propto
t^{(8+k+4m)/[4(m+1)]},   \phantom{sssssssss}  \nu_{c}<\nu<\nu_{a}, \\
\displaystyle(\frac{\nu}{\nu_{c}})^{-1/2}F_{\nu,\rm{max}}\propto
t^{(8-3k-4m)/[4(m+1)]},   \phantom{sssssssssssssss}  \nu_{a}<\nu<\nu_{m},\\
\displaystyle(\frac{\nu}{\nu_{m}})^{-p/2}(\frac{\nu_{m}}{\nu_{c}})^{-1/2}F_{\nu,\rm{max}}\propto
t^{[8-2k-p(4m+k)]/[4(m+1)]},   \phantom{ss}\,  \nu_{m}<\nu,\\
 \end{array} \right.
\label{eqn:fc:syn-spectrum-1}
\end{equation}
while for $t_{ac}<t<t_{cm}$, the flux density is
\begin{equation}
F_{\nu}=\left\{
\begin{array}{l}
\displaystyle(\frac{\nu}{\nu_{a}})^{2}(\frac{\nu_{a}}{\nu_{c}})^{1/3}F_{\nu,\rm{max}}\propto
t^{(1+k+m)/(m+1)},   \phantom{sssssssssssssss}  \nu<\nu_{a}, \\
\displaystyle(\frac{\nu}{\nu_{c}})^{1/3}F_{\nu,\rm{max}}\propto
t^{(11-6k-3m)/[3(m+1)]},   \phantom{ssssssssssssssss}  \nu_{a}<\nu<\nu_{c}, \\
\displaystyle(\frac{\nu}{\nu_{c}})^{-1/2}F_{\nu,\rm{max}}\propto
t^{(8-3k-4m)/[4(m+1)]},   \,\phantom{sssssssssssssss}  \nu_{c}<\nu<\nu_{m},\\
\displaystyle(\frac{\nu}{\nu_{m}})^{-p/2}(\frac{\nu_{m}}{\nu_{c}})^{-1/2}F_{\nu,\rm{max}}\propto
t^{[8-2k-p(4m+k)]/[4(m+1)]},   \,\phantom{ss}\,  \nu_{m}<\nu.\\
 \end{array} \right.
\label{eqn:fc:syn-spectrum-2}
\end{equation}

\subsection{Constraint on the IC component in an X-ray afterglow}
The synchrotron-self-Compton (SSC) spectrum is featured by the
characteristic IC frequencies, i.e., $\nu_{a}^{\rm{IC}}\approx
2\gamma_{c}^2\nu_{a}$, $\nu_{m}^{\rm{IC}}\approx
2\gamma_{e,\rm{min}}^2\nu_{m}$ and $\nu_{c}^{\rm{IC}}\approx
2\gamma_{c}^2\nu_{c}$. The IC frequency $\nu_{a}^{\rm{IC}}$ can be
directly determined by
\begin{equation}
\nu_{a}^{\rm{IC}}=\rm{min}\{\it{\nu_{a,<}^{\rm{IC}},\nu_{a,>}^{\rm{IC}}}\},
\label{eqn:fc:nu_a-ic}
\end{equation}
where $\nu_{a,<}^{\rm{IC}}\approx 2\gamma_{c}^2\nu_{a,<}$ and
$\nu_{a,>}^{\rm{IC}}\approx 2\gamma_{c}^2\nu_{a,>}$. Inserting
equations (\ref{eqn:fc:gc}), (\ref{eqn:gamma_cm-2}),
(\ref{eqn:fc:nu_a<}) and (\ref{eqn:fc:nu_a>}) into the above
equation, we obtain
\begin{equation}
\nu_{a,<}^{\rm{IC}}=\left\{ \begin{array}{l}
2.24\times10^{16}(1+z)^{-1}\epsilon_{e,-0.5}^{-5/4}\epsilon_{B,-2.5}^{-11/20}\zeta_{1/6}^{-1/2}E_{cm,53}^{-3/10}n^{-2/5}(\displaystyle\frac{t}{t_{cm}})^{-(5-2\epsilon)/[5(4-\epsilon)]}\,\,\rm{Hz}, \phantom{ss}  \rm{ISM,} \\
3.01\times10^{16}(1+z)^{-1}\epsilon_{e,-0.5}^{-13/20}\epsilon_{B,-2.5}^{-7/20}\zeta_{1/6}^{-1/10}E_{cm,53}^{1/10}A_{\ast}^{-2/5}(\displaystyle\frac{t}{t_{cm}})^{-(1+30\epsilon)/[5(2-\epsilon)]}\,\,\rm{Hz},  \;  \rm{wind,}\\
 \end{array} \right.
 \label{eqn:fc:nu_a-ic<}
\end{equation}
while the expression for $\nu_{a,>}^{\rm{IC}}$ is
\begin{equation}
\nu_{a,>}^{\rm{IC}}=\left\{ \begin{array}{l}
1.19\times10^{18}(1+z)^{-1}\epsilon_{e,-0.5}^{-23/12}\epsilon_{B,-2.5}^{-13/12}\zeta_{1/6}^{-1/2}E_{cm,53}^{-5/6}n^{-4/3}(\displaystyle\frac{t}{t_{cm}})^{-(1-2\epsilon)/(4-\epsilon)}\,\,\rm{Hz}, \phantom{s}  \rm{ISM,} \\
4.27\times10^{17}(1+z)^{-1}\epsilon_{e,-0.5}^{1/12}\epsilon_{B,-2.5}^{-5/12}\zeta_{1/6}^{5/6}E_{cm,53}^{1/2}A_{\ast}^{-4/3}(\displaystyle\frac{t}{t_{cm}})^{(5-14\epsilon)/[3(2-\epsilon)]}\,\,\rm{Hz},  \phantom{ss}  \rm{wind.}\\
 \end{array} \right.
  \label{eqn:fc:nu_a-ic>}
\end{equation}
As we can see, $\nu_{a}^{\rm{IC}}$ is below the X-ray frequency
$\nu\sim10^{18}$ Hz for typical parameters in most times during
the fast-cooling phase. For simplicity, we do not consider this
frequency for our estimation of the IC component in the X-ray
light curve.

The SSC frequency $\nu_{m}^{\rm{IC}}$ equals to
$\nu_{c}^{\rm{IC}}$ when $t=t_{cm}$, i.e.
$\nu_{cm}^{\rm{IC}}\equiv 2\gamma_{e,cm}^2\nu_{cm}$. According to
equations (\ref{eqn:gamma_cm-1}) and (\ref{eqn:nu_0-1}), we obtain
\begin{eqnarray}
\nu_{cm}^{\rm{IC}}=&&\displaystyle\frac{1}{128}(\frac{3}{2\pi})^{5/4}\frac{m_{e}c^3}{e^2(1+z)}(\frac{2m_{p}}{3m_{e}})^{(5k-9)/2}\epsilon_{e}^{(15k-22)/8}\epsilon_{B}^{(5k-14)/8}
            \zeta_{1/6}^{(5k-2)/4}\nonumber\\
        &&\times(A\sigma_{\rm{T}}^{(3-k)/2})^{-5/2}[\frac{(3-k)E_{cm}}{4\pi m_{p}c^{2}}]^{(5k-6)/4}, \label{eqn:nu_0-ic-1}
\end{eqnarray}
which can be further reduced numerically as
\begin{equation}
\nu_{cm}^{\rm{IC}}=\left\{ \begin{array}{l}
1.72\times 10^{20}(1+z)^{-1}\epsilon_{e,-0.5}^{-11/4}\epsilon_{B,-2.5}^{-7/4}\zeta_{1/6}^{-1/2}E_{cm,53}^{-3/2}n^{-5/2}\,\,\rm{Hz}, \phantom{ss}  \rm{ISM,} \\
1.17\times 10^{19}(1+z)^{-1}\epsilon_{e,-0.5}\epsilon_{B,-2.5}^{-1/2}\zeta_{1/6}^{2}E_{cm,53}A_{\ast}^{-5/2}\,\,\rm{Hz},  \phantom{sss}  \rm{wind.}\\
 \end{array} \right.
 \label{eqn:nu_0-ic-2}
\end{equation}
The characteristic SSC frequencies $\nu_{c}^{\rm{IC}}$ and
$\nu_{m}^{\rm{IC}}$ evolve with time as
\begin{equation}
\nu_{c}^{\rm{IC}}=\nu_{cm}^{\rm{IC}}(\frac{t}{t_{cm}})^{(2m+7k-8)/[2(m+1)]},\phantom{ss}
\nu_{m}^{\rm{IC}}=\nu_{cm}^{\rm{IC}}(\frac{t}{t_{cm}})^{-(6m+k)/[2(m+1)]}.
\end{equation}
The peak flux density of the SSC spectrum,
$F_{\nu,\rm{max}}^{\rm{IC}}$, is roughly the product of the peak
flux density of the synchrotron spectrum by the Thomson optical
depth. Considering some numerical factors of order unity, the
exact expression is (Sari \& Esin 2001)
\begin{equation}
F_{\nu,\rm{max}}^{\rm{IC}}\approx\frac{28}{45}x_{0}(\sigma_{\rm{T}}Rn)F_{\nu,\rm{max}}\propto
t^{(8-5k-2m)/[2(m+1)]},
\end{equation}
where $x_{0}\approx0.5$. The inverse Compton spectrum above
$\nu_{a}^{\rm{IC}}$ is similar to the synchrotron one, which can
be approximated by several power law segments, i.e.
\begin{equation}
F_{\nu}^{\rm{IC}}=\left\{
\begin{array}{l}
\displaystyle(\frac{\nu}{\nu_{c}^{\rm{IC}}})^{1/3}F_{\nu,\rm{max}}^{\rm{IC}}\propto
t^{(16-11k-4m)/[3(m+1)]},   \phantom{sssssssssssssssss}  \nu<\nu_{c}^{\rm{IC}}, \\
\displaystyle(\frac{\nu}{\nu_{c}^{\rm{IC}}})^{-1/2}F_{\nu,\rm{max}}^{\rm{IC}}\propto
t^{(8-3k-2m)/[4(m+1)]},   \phantom{sssssssssssssssss}  \nu_{c}^{\rm{IC}}<\nu<\nu_{m}^{\rm{IC}},\\
\displaystyle(\frac{\nu}{\nu_{m}^{\rm{IC}}})^{-p/2}(\frac{\nu_{m}^{\rm{IC}}}{\nu_{c}^{\rm{IC}}})^{-1/2}F_{\nu,\rm{max}}^{\rm{IC}}\propto
t^{[8-2k+4m-p(6m+k)]/[4(m+1)]},   \phantom{s}\,  \nu_{m}^{\rm{IC}}<\nu,\\
 \end{array} \right.
\label{eqn:fc:SSC-spectrum}
\end{equation}
where we have neglected the logarithmic term for
$\nu>\nu_{m}^{\rm{IC}}$.

The SSC flux density begins to dominate over that of the
synchrotron radiation in the overall synchrotron $+$ SSC spectrum
at the crossing point, which corresponds to
$\nu_{\times}^{\rm{IC}}$ (Sari \& Esin 2001). Using equation
(\ref{eqn:fc:SSC-spectrum}) and the standard synchrotron spectrum,
and assuming $\nu_{\times}^{\rm{IC}}>\nu_{m}>\nu_{c}$, we obtain
the crossing point frequency for two cases,
$\nu_{\times}^{\rm{IC}}<\nu_{c}^{\rm{IC}}$ and
$\nu_{c}^{\rm{IC}}<\nu_{\times}^{\rm{IC}}<\nu_{m}^{\rm{IC}}$, i.e.
\begin{equation}
\nu_{\times}^{\rm{IC}}=\left\{
\begin{array}{l}
\nu_{\times,<}^{\rm{IC}}\equiv\nu_{c}^{\rm{IC}}[c_1\displaystyle\frac{\epsilon_{B}}{\epsilon_{e}}(\frac{\gamma_{e,\rm{min}}}{\gamma_{c}})^{3p-2}(2\gamma_{c}\gamma_{e,\rm{min}})^{2-p}]^{3/(2+3p)}, \phantom{s} \rm{if} \phantom{ss} \nu_{\times}^{\rm{IC}}<\nu_{c}^{\rm{IC}}, \\
\nu_{\times,>}^{\rm{IC}}\equiv\nu_{c}^{\rm{IC}}[c_1\displaystyle\frac{\epsilon_{B}}{\epsilon_{e}}(\frac{\gamma_{e,\rm{min}}}{\gamma_{c}})^{3p-2}(2\gamma_{c}\gamma_{e,\rm{min}})^{2-p}]^{1/(p-1)}, \phantom{ss} \rm{if} \phantom{ss} \nu_{c}^{\rm{IC}}<\nu_{\times}^{\rm{IC}}<\nu_{m}^{\rm{IC}}, \\
 \end{array} \right.
\label{eqn:fc:nu-ic-1}
\end{equation}
where the coefficient is
$c_1=\displaystyle\frac{225(1-\epsilon)^2(p-1)^2}{49x_{0}^2(4-k-\epsilon)^2(p-2)^2}$.
To derive the above equation we have used the relation
\begin{equation}
\gamma_{c}\gamma_{e,\rm{min}}=\displaystyle\frac{3(4-k-\epsilon)(p-2)}{8(1-\epsilon)(p-1)(1+Y)}\frac{\epsilon_{e}}{\epsilon_{B}}\frac{1}{\sigma_{\rm{T}}nR}.
\end{equation}
Since $1/(p-1)$ is always larger than
$3/(2+3p)$, one can determine
$\nu_{\times}^{\rm{IC}}$ directly by
\begin{equation}
\nu_{\times}^{\rm{IC}}=\rm{max}\{\nu_{\times,<}^{\rm{IC}},\nu_{\times,>}^{\rm{IC}}\},
\label{eqn:fc:nu-ic-2}
\end{equation}
without judging whether $\nu_{\times}^{\rm{IC}}<\nu_{c}^{\rm{IC}}$
or not.

We have calculated numerically the temporal evolution of
$\nu_{\times,<}^{\rm{IC}}$ and $\nu_{\times,>}^{\rm{IC}}$ for
typical physical parameters. {\em In the ISM case}, the expression
for $\nu_{\times,<}^{\rm{IC}}$ is
\begin{equation}
\nu_{\times,<}^{\rm{IC}}=\left\{
\begin{array}{l}
3.5\times 10^{19}(1+z)^{-1}\epsilon_{e,-0.5}^{-3.08}\epsilon_{B,-2.5}^{-1.35}E_{cm,53}^{-1.47}n^{-2.43}(\displaystyle\frac{t}{t_{cm}})^{(10\epsilon-17.8)/[4.3(4-\epsilon)]}\,{\rm{Hz}},\, p=2.2, \\
9.1\times 10^{18}(1+z)^{-1}\epsilon_{e,-0.5}^{-3.04}\epsilon_{B,-2.5}^{-1.33}E_{cm,53}^{-1.43}n^{-2.37}(\displaystyle\frac{t}{t_{cm}})^{(5\epsilon-9.8)/[2.3(4-\epsilon)]}\,{\rm{Hz}},\phantom{ss} p=2.4, \\
 \end{array} \right.
 \label{eqn:fc:ism:nu<-ic}
\end{equation}
while the expression for $\nu_{\times,>}^{\rm{IC}}$ is
\begin{equation}
\nu_{\times,>}^{\rm{IC}}=\left\{
\begin{array}{l}
3.9\times 10^{18}(1+z)^{-1}\epsilon_{e,-0.5}^{-3.54}\epsilon_{B,-2.5}^{-0.79}E_{cm,53}^{-1.42}n^{-2.33}(\displaystyle\frac{t}{t_{cm}})^{-8.5/(4-\epsilon)}\,{\rm{Hz}},\phantom{ss} p=2.2, \\
3.0\times 10^{17}(1+z)^{-1}\epsilon_{e,-0.5}^{-3.39}\epsilon_{B,-2.5}^{-0.82}E_{cm,53}^{-1.36}n^{-2.21}(\displaystyle\frac{t}{t_{cm}})^{-57/[7(4-\epsilon)]}\,{\rm{Hz}},\phantom{s} p=2.4. \\
 \end{array} \right.
  \label{eqn:fc:ism:nu>-ic}
\end{equation}
The crossing point frequency $\nu_{\times}^{\rm{IC}}$ decreases
throughout the fast cooling phase in the ISM case. The transition
from the initial $\nu_{\times}^{\rm{IC}}=\nu_{\times,>}^{\rm{IC}}$
to the late $\nu_{\times}^{\rm{IC}}=\nu_{\times,<}^{\rm{IC}}$
happens when $\nu_{\times}^{\rm{IC}}=\nu_{c}^{\rm{IC}}$, i.e. at
\begin{equation}
t_{\times,c}^{\rm{IC}}=\left\{
\begin{array}{l}
0.24\epsilon_{e,-1}^{-0.39}\epsilon_{B,-2.5}^{0.48}E_{cm,53}^{0.04}n^{0.08}t_{cm},\,\phantom{ss} \rm{if}\,\,\epsilon=0.1, \\
0.20\epsilon_{e,-0.5}^{-0.33}\epsilon_{B,-2.5}^{0.40}E_{cm,53}^{0.04}n^{0.07}t_{cm},\phantom{ss} \rm{if}\,\,\epsilon=0.32, \\
 \end{array} \right.
\end{equation}
for $p=2.2$, and
\begin{equation}
t_{\times,c}^{\rm{IC}}=\left\{
\begin{array}{l}
0.057\epsilon_{e,-1}^{-0.33}\epsilon_{B,-2.5}^{0.49}E_{cm,53}^{0.07}n^{0.15}t_{cm},\,\phantom{ss} \rm{if}\,\,\epsilon=0.1, \\
0.064\epsilon_{e,-0.5}^{-0.28}\epsilon_{B,-2.5}^{0.41}E_{cm,53}^{0.06}n^{0.13}t_{cm},\phantom{ss} \rm{if}\,\,\epsilon=0.32, \\
 \end{array} \right.
\end{equation}
for $p=2.4$. The condition for the appearance of the IC component
in the soft X-ray afterglow is that $\nu_{\times}^{\rm{IC}}$ at
$t_{cm}$ must be much less than $\nu=10^{18}\nu_{18}$ Hz, which
leads to the lower limit on the ambient density $n$. Using
equation (\ref{eqn:fc:ism:nu<-ic}), we obtain the lower limit of
$n$ as
\begin{equation}
n\gtrsim\left\{
\begin{array}{l}
4.3(1+z)^{-0.41}\nu_{18}^{-0.41}\epsilon_{e,-0.5}^{-1.27}\epsilon_{B,-2.5}^{-0.56}E_{cm,53}^{-0.60}\,{\rm{cm}^{-3}},\phantom{ss} p=2.2, \\
2.5(1+z)^{-0.42}\nu_{18}^{-0.42}\epsilon_{e,-0.5}^{-1.28}\epsilon_{B,-2.5}^{-0.56}E_{cm,53}^{-0.60}\,{\rm{cm}^{-3}},\phantom{ss} p=2.4. \\
 \end{array} \right.
\end{equation}
The lower limit of $n$ for the emergence of IC component in the
X-ray afterglow in the fast cooling phase is typically
$\sim1$--$10$ cm$^{-3}$ (Panaitescu \& Kumar 2000).

{\em In the stellar wind case}, the expression for
$\nu_{\times,<}^{\rm{IC}}$ is
\begin{equation}
\nu_{\times,<}^{\rm{IC}}=\left\{
\begin{array}{l}
4.5\times 10^{18}(1+z)^{-1}\epsilon_{e,-0.5}^{0.56}\epsilon_{B,-2.5}^{-0.13}E_{cm,53}^{0.97}A_{\ast}^{-2.43}(\displaystyle\frac{t}{t_{cm}})^{(3.1-5.7\epsilon)/[4.3(2-\epsilon)]}\,{\rm{Hz}},\phantom{s} p=2.2, \\
1.4\times 10^{18}(1+z)^{-1}\epsilon_{e,-0.5}^{0.48}\epsilon_{B,-2.5}^{-0.12}E_{cm,53}^{0.93}A_{\ast}^{-2.36}(\displaystyle\frac{t}{t_{cm}})^{(2-9\epsilon)/[8.6(2-\epsilon)]}\,{\rm{Hz}},\,\phantom{sss} p=2.4, \\
 \end{array} \right.
  \label{eqn:fc:wind:nu<-ic}
\end{equation}
while the expression for $\nu_{\times,>}^{\rm{IC}}$ is
\begin{equation}
\nu_{\times,>}^{\rm{IC}}=\left\{
\begin{array}{l}
1.2\times 10^{18}(1+z)^{-1}\epsilon_{e,-0.5}^{-0.04}\epsilon_{B,-2.5}^{0.38}E_{cm,53}^{0.92}A_{\ast}^{-2.33}(\displaystyle\frac{t}{t_{cm}})^{(6\epsilon-23)/[6(2-\epsilon)]}\,{\rm{Hz}},\phantom{ss} p=2.2, \\
1.1\times 10^{17}(1+z)^{-1}\epsilon_{e,-0.5}^{-0.07}\epsilon_{B,-2.5}^{0.29}E_{cm,53}^{0.86}A_{\ast}^{-2.21}(\displaystyle\frac{t}{t_{cm}})^{(7\epsilon-26)/[7(2-\epsilon)]}\,{\rm{Hz}},\phantom{ss} p=2.4. \\
 \end{array} \right.
   \label{eqn:fc:wind:nu>-ic}
\end{equation}
The crossing point frequency decreases with
$\nu_{\times}^{\rm{IC}}=\nu_{\times,>}^{\rm{IC}}$ initially.
However, the temporal behavior of the crossing point frequency at
late times, $\nu_{\times}^{\rm{IC}}=\nu_{\times,<}^{\rm{IC}}$,
depends on $\epsilon$. The transition time when
$\nu_{\times,<}^{\rm{IC}}=\nu_{\times,>}^{\rm{IC}}=\nu_{c}^{\rm{IC}}$,
is
\begin{equation}
t_{\times,c}^{\rm{IC}}=\left\{
\begin{array}{l}
0.75\epsilon_{e,-1}^{-0.26}\epsilon_{B,-2.5}^{0.22}E_{cm,53}^{-0.02}A_{\ast}^{0.04}t_{cm},\,\phantom{ss} \rm{if}\,\,\epsilon=0.1, \\
0.56\epsilon_{e,-0.5}^{-0.26}\epsilon_{B,-2.5}^{0.22}E_{cm,53}^{-0.02}A_{\ast}^{0.04}t_{cm},\phantom{ss} \rm{if}\,\,\epsilon=0.32, \\
 \end{array} \right.
\end{equation}
for $p=2.2$, and
\begin{equation}
t_{\times,c}^{\rm{IC}}=\left\{
\begin{array}{l}
0.38\epsilon_{e,-1}^{-0.28}\epsilon_{B,-2.5}^{0.21}E_{cm,53}^{-0.04}A_{\ast}^{0.08}t_{cm},\,\phantom{ss} \rm{if}\,\,\epsilon=0.1, \\
0.27\epsilon_{e,-0.5}^{-0.28}\epsilon_{B,-2.5}^{0.21}E_{cm,53}^{-0.04}A_{\ast}^{0.08}t_{cm},\phantom{ss} \rm{if}\,\,\epsilon=0.32, \\
 \end{array} \right.
\end{equation}
for $p=2.4$. After this time, the crossing point frequency
$\nu_{\times}^{\rm{IC}}=\nu_{\times,<}^{\rm{IC}}$ will increase
for $\epsilon<31/57$ ($2/9$), or continue to decrease for
$\epsilon>31/57$ ($2/9$) for $p=2.2$ ($p=2.4$). The emergence of
the IC component in the soft X-ray afterglow requires that the
minimum of $\nu_{\times}^{\rm{IC}}$ during the fast
cooling phase is below $\nu=10^{18}\nu_{18}$ Hz, which leads to
a lower limit of $A_{\ast}$ as
\begin{equation}
A_{\ast}\gtrsim\left\{
\begin{array}{l}
1.88(1+z)^{-0.41}\nu_{18}^{-0.41}\epsilon_{e,-0.5}^{0.20}\epsilon_{B,-2.5}^{-0.02}E_{cm,53}^{0.40},\phantom{ss} \rm{if}\,\,\epsilon<31/57, \\
2.18(1+z)^{-0.41}\nu_{18}^{-0.41}\epsilon_{e,-0.2}^{0.23}\epsilon_{B,-2.5}^{-0.05}E_{cm,53}^{0.40},\phantom{ss} \rm{if}\,\,\epsilon>31/57, \\
 \end{array} \right.
\end{equation}
for $p=2.2$, and
\begin{equation}
A_{\ast}\gtrsim\left\{
\begin{array}{l}
0.94(1+z)^{-0.43}\nu_{18}^{-0.43}\epsilon_{e,-1}^{0.20}\epsilon_{B,-2.5}^{-0.05}E_{cm,53}^{0.40},\;\;\phantom{ss} \rm{if}\,\,\epsilon<2/9, \\
1.15(1+z)^{-0.42}\nu_{18}^{-0.42}\epsilon_{e,-0.5}^{0.20}\epsilon_{B,-2.5}^{-0.05}E_{cm,53}^{0.39},\phantom{ss} \rm{if}\,\,\epsilon>2/9, \\
 \end{array} \right.
\end{equation}
for $p=2.4$. The above constraint on $A_{\ast}$ is insensitive to
the other physical parameters. This implies that the contribution
of the IC component to the X-ray afterglow can be neglected during
the fast-cooling phase for typical values of $A_{\ast}\lesssim1$ as
indicated from the observations of Wolf-Rayet stars and the
fittings to some GRB afterglows (Chevalier \& Li 1999; Chevalier, Li, \&
Fransson 2004).

\section{The late slow-cooling phase}
\label{sec:slowcooling} The hydrodynamic evolution of the blast
wave in the slow cooling phase can be approximated as that in the
early fast cooling phase. This approximation is validated by the
fact that the radiation efficiency of the blast wave in the slow
cooling phase evolves as,
\begin{equation}
\epsilon=\epsilon_{e}(\displaystyle\frac{\nu_{m}}{\nu_{c}})^{(p-2)/2},
\end{equation}
which decreases very slowly with time as long as the electron
power law index $p$ does not deviate far from $2$, e.g. $p\sim
2.2$ as expected in the relativistic shock acceleration theory
(Achterberg et al. 2001). This fact will prolong the
semi-radiative phase by at least two orders of magnitude in time
than $t_{cm}$ for typical $\epsilon_{e}\sim 1/3$ and
$p=2.2$\footnotemark\footnotetext{\label{foot:slowcooling}The
hydrodynamics will deviate significantly from that of the constant
radiation efficiency $\epsilon=\epsilon_{e}$ in the very late
times of the slow cooling phase when $\epsilon<e^{-1}\epsilon_{e}$
($e\approx2.71828$). This happens when $t\gtrsim 1300 t_{cm}$ ($90
t_{cm}$) for the ISM (wind) case for $p=2.2$ and
$\epsilon_{e}=0.32$ (see equations \ref{eqn:fc:spectrum} and
\ref{eqn:sc:spectrum}). It will happen even later if we adopt the
adiabatic relations for $\nu_{c}$ and $\nu_{m}$.}. Hereafter we
assume $\epsilon\approx\epsilon_{e}$ for
$t>t_{cm}$. The time when $\epsilon\ll\epsilon_{e}$ happens is expected to
be very late, when the Lorentz factor of GRB conical ejecta has
already dropped below the inverse of the initial opening
angle and the resulting light curves deviate from the
spherical-like ones, which is beyond the scope of this paper.

\subsection{Properties of the synchrotron radiation}
The temporal behaviors of the typical frequency $\nu_{m}$ and peak
flux density $F_{\nu,\rm{max}}$ in the slow cooling phase are the
same as in the fast cooling phase. However, the Compton parameter
in the slow cooling case is no longer a constant but evolves as
$Y=\displaystyle\sqrt{\frac{\epsilon_{e}}{\epsilon_{B}}}(\displaystyle\frac{\nu_{m}}{\nu_{c}})^{(p-2)/4}$.
Since the cooling Lorentz factor and the cooling frequency behave
as $\gamma_{c}\propto(1+Y)^{-1}$ and
$\nu_{c}\propto\gamma_{c}^{2}$, we obtain
\begin{eqnarray}
\gamma_{c}&=&\displaystyle\gamma_{e,cm}(\frac{t}{t_{cm}})^{(mp+4k-4)/[2(4-p)(m+1)]},\nonumber\\
\nu_{c}&=&\displaystyle\nu_{cm}(\frac{t}{t_{cm}})^{[6k-8+(p-2)(4m+k)]/[2(4-p)(m+1)]},\nonumber\\
Y&=&\displaystyle\sqrt{\frac{\epsilon_{e}}{\epsilon_{B}}}(\frac{t}{t_{cm}})^{-[(p-2)(m+k-1)]/[(4-p)(m+1)]}.
\label{eqn:sc:spectrum}
\end{eqnarray}

The SSA frequency in the slow cooling phase is
\begin{equation}
\nu_{a}=\left\{ \begin{array}{l}
\nu_{a,<}\equiv\nu_{m}\displaystyle[\frac{c_{0}(p-1)}{(3-k)}\frac{enR}{B\gamma_{e,\rm{min}}^{5}}]^{3/5}, \phantom{sssss} \rm{if} \phantom{ss} \nu_{a}<\nu_{m}, \\
\nu_{a,>}\equiv\nu_{m}\displaystyle[\frac{c_{0}(p-1)}{(3-k)}\frac{enR}{B\gamma_{e,\rm{min}}^{5}}]^{2/(p+4)}, \phantom{ss} \rm{if} \phantom{ss} \nu_{m}<\nu_{a}<\nu_{c},\\
 \end{array} \right.
 \label{eqn:sc:nu_a}
\end{equation}
which can be determined by
$\nu_{a}=\rm{min}\{\it{\nu_{a,<},\nu_{a,>}}\}$ without judging
whether $\nu_{a}<\nu_{m}$ or not. The case for $\nu_{a}>\nu_{c}$
can be neglected. The numerical expression for $\nu_{a,<}$ is
\begin{equation}
\nu_{a,<}=\left\{ \begin{array}{l}
3.72\displaystyle\kappa_{p}^{3/5}(1+z)^{-1}\epsilon_{e,-0.5}^{-1}\epsilon_{B,-2.5}^{1/5}\zeta_{1/6}^{-1}E_{cm,53}^{1/5}n^{3/5}(\displaystyle\frac{t}{t_{cm}})^{-3\epsilon/[5(4-\epsilon)]}\,\,\rm{GHz}, \,\phantom{sssss}  \rm{ISM,} \\
16.4\displaystyle\kappa_{p}^{3/5}(1+z)^{-1}\epsilon_{e,-0.5}^{-19/10}\epsilon_{B,-2.5}^{-1/10}\zeta_{1/6}^{-8/5}E_{cm,53}^{-2/5}A_{\ast}^{3/5}(\displaystyle\frac{t}{t_{cm}})^{(5\epsilon-6)/[5(2-\epsilon)]}\,\,\rm{GHz},  \phantom{s}  \rm{wind,}\\
 \end{array} \right.
 \label{eqn:sc:nu_a<}
\end{equation}
where $\kappa_{p}=(p-1)(p+2)/(p+2/3)$, while
the expression for $\nu_{a,>}$ is
\begin{equation}
\nu_{a,>}=\left\{ \begin{array}{l}
3.13\times10^{11}(1+z)^{-1}\epsilon_{e,-0.5}^{-1.69}\epsilon_{B,-2.5}^{-0.35}E_{cm,53}^{-0.35}n^{-0.37}(\displaystyle\frac{t}{t_{cm}})^{-(8.6+\epsilon)/[3.1(4-\epsilon)]}\,\,\rm{Hz}, \phantom{ssss}  \rm{ISM,} \\
3.46\times10^{11}(1+z)^{-1}\epsilon_{e,-0.5}^{-1.14}\epsilon_{B,-2.5}^{-0.17}E_{cm,53}^{0.02}A_{\ast}^{-0.37}(\displaystyle\frac{t}{t_{cm}})^{-(6.3-3.1\epsilon)/[3.1(2-\epsilon)]}\,\,\rm{Hz},  \phantom{ss}  \rm{wind,}\\
 \end{array} \right.
  \label{eqn:sc:nu_a>-1}
\end{equation}
for $p=2.2$, and
\begin{equation}
\nu_{a,>}=\left\{ \begin{array}{l}
2.20\times10^{11}(1+z)^{-1}\epsilon_{e,-0.5}^{-1.72}\epsilon_{B,-2.5}^{-0.38}E_{cm,53}^{-0.38}n^{-0.41}(\displaystyle\frac{t}{t_{cm}})^{-(9.2+\epsilon)/[3.2(4-\epsilon)]}\,\,\rm{Hz}, \phantom{ssss}  \rm{ISM,} \\
2.89\times10^{11}(1+z)^{-1}\epsilon_{e,-0.5}^{-1.10}\epsilon_{B,-2.5}^{-0.17}E_{cm,53}^{0.03}A_{\ast}^{-0.41}(\displaystyle\frac{t}{t_{cm}})^{-(3.3-1.6\epsilon)/[1.6(2-\epsilon)]}\,\,\rm{Hz},  \phantom{ss}  \rm{wind,}\\
 \end{array} \right.
  \label{eqn:sc:nu_a>-2}
\end{equation}
for $p=2.4$. In the ISM case, the transition from the earlier
$\nu_{a}=\nu_{a,<}$ to the later $\nu_{a}=\nu_{a,>}$ happens when
$\nu_{a}=\nu_{m}$, at
\begin{equation}
t_{am}=\left\{ \begin{array}{l}
1.3\times10^{3}\displaystyle\kappa_{p}^{0.39}\epsilon_{e,-1}^{-0.98}\epsilon_{B,-2.5}^{-0.79}E_{cm,53}^{-0.79}n^{-1.38}t_{cm}, \phantom{ss}  \rm{if} \phantom{ss} \epsilon=0.1, \\
3.4\times10^{2}\displaystyle\kappa_{p}^{0.38}\epsilon_{e,-0.5}^{-0.95}\epsilon_{B,-2.5}^{-0.76}E_{cm,53}^{-0.76}n^{-1.33}t_{cm}, \phantom{ss}  \rm{if} \phantom{ss} \epsilon=0.32,\\
\end{array} \right.
\end{equation}
which indicates that the time when $\nu_{a}=\nu_{a,>}$ is very
late, unless the medium is very dense. In the stellar wind case,
the transition from $\nu_{a}=\nu_{a,<}$ to $\nu_{a}=\nu_{a,>}$
takes place when $\nu_{a}=\nu_{m}$, at
\begin{equation}
t_{am}=\left\{ \begin{array}{l}
93.7\displaystyle\kappa_{p}^{0.63}\epsilon_{e,-1}^{1.74}\epsilon_{B,-2.5}^{-0.16}\zeta_{1/6}^{2.22}E_{cm,53}^{0.95}A_{\ast}^{-2.22}t_{cm}, \phantom{ssss}  \rm{if} \phantom{ss} \epsilon=0.1, \\
330.5\displaystyle\kappa_{p}^{0.56}\epsilon_{e,-0.5}^{1.54}\epsilon_{B,-2.5}^{-0.14}\zeta_{1/6}^{1.96}E_{cm,53}^{0.84}A_{\ast}^{-1.96}t_{cm}, \phantom{ss}  \rm{if} \phantom{ss} \epsilon=0.32,\\
\end{array} \right.
\end{equation}
which also indicates that the time when $\nu_{a}=\nu_{a,>}$ is
very late for the stellar wind case. Therefore we neglect
the case of $t>t_{am}$ below.

The flux density at the observed frequency $\nu$ from the
synchrotron component for $t_{cm}<t<t_{am}$ is
\begin{equation}
F_{\nu}=\left\{
\begin{array}{l}
\displaystyle(\frac{\nu}{\nu_{a}})^{2}(\frac{\nu_{a}}{\nu_{m}})^{1/3}F_{\nu,\rm{max}}\propto
t^{2/(m+1)},   \phantom{sssssssssssssssssssssssssssssssssssssssssss}  \nu<\nu_{a}, \\
\displaystyle(\frac{\nu}{\nu_{m}})^{1/3}F_{\nu,\rm{max}}\propto
t^{(9-4k-m)/[3(m+1)]},   \phantom{sssssssssssssssssssssssssssssssssssssss}  \nu_{a}<\nu<\nu_{m}, \\
\displaystyle(\frac{\nu}{\nu_{m}})^{-(p-1)/2}F_{\nu,\rm{max}}\propto
t^{(12-5k-p(4m+k))/[4(m+1)]},   \,\phantom{sssssssssssssssssssssssssssss}  \nu_{m}<\nu<\nu_{c},\\
\displaystyle(\frac{\nu}{\nu_{c}})^{-p/2}(\frac{\nu_{c}}{\nu_{m}})^{-(p-1)/2}F_{\nu,\rm{max}}\propto
t^{\{(4-p)[12-6k-4m-p(4m+k)]+8(k+m-1)\}/[4(4-p)(m+1)]},   \phantom{s}  \nu_{c}<\nu.\\
 \end{array} \right.
\label{eqn:sc:syn-spectrum}
\end{equation}

\subsection{Constraint on the IC component in an X-ray afterglow}
The temporal behaviors of the typical SSC frequency
$\nu_{m}^{\rm{IC}}\approx 2\gamma_{e,\rm{min}}^2\nu_{m}$ and peak
flux density $F_{\nu,\rm{max}}^{\rm{IC}}$ in the IC spectrum are
the same in the slow cooling phase as in the fast cooling phase.
The IC frequency corresponding to $\nu_{c}$ is
\begin{equation}
\nu_{c}^{\rm{IC}} \approx 2\gamma_{c}^2\nu_{c}=\displaystyle\nu_{cm}^{\rm{IC}}(\frac{t}{t_{cm}})^{(6mp+pk+12k-8m-16)/[2(4-p)(m+1)]}.
\end{equation}
The IC frequency $\nu_{a}^{\rm{IC}}$ can be directly determined by
$\nu_{a}^{\rm{IC}}=\rm{min}\{\it{\nu_{a,<}^{\rm{IC}},\nu_{a,>}^{\rm{IC}}}\}$,
where $\nu_{a,<}^{\rm{IC}}\approx 2\gamma_{e,\rm{min}}^2\nu_{a,<}$
and $\nu_{a,>}^{\rm{IC}}\approx 2\gamma_{e,\rm{min}}^2\nu_{a,>}$.
Inserting equations (\ref{eqn:gm}), (\ref{eqn:gamma_cm-2}), and
(\ref{eqn:sc:nu_a<}) -- (\ref{eqn:sc:nu_a>-2}) into the above
equation, we obtain
\begin{equation}
\nu_{a,<}^{\rm{IC}}=\left\{ \begin{array}{l}
1.75\times10^{16}\displaystyle\kappa_{p}^{3/5}(1+z)^{-1}\epsilon_{e,-0.5}^{-5/4}\epsilon_{B,-2.5}^{-11/20}\zeta_{1/6}^{-1/2}E_{cm,53}^{-3/10}n^{-2/5}(\displaystyle\frac{t}{t_{cm}})^{-(15+3\epsilon)/[5(4-\epsilon)]}\,\,\rm{Hz},  \phantom{s}\rm{ISM,} \\
2.36\times10^{16}\displaystyle\kappa_{p}^{3/5}(1+z)^{-1}\epsilon_{e,-0.5}^{-13/20}\epsilon_{B,-2.5}^{-7/20}\zeta_{1/6}^{-1/10}E_{cm,53}^{1/10}A_{\ast}^{-2/5}(\displaystyle\frac{t}{t_{cm}})^{-(11-5\epsilon)/[5(2-\epsilon)]}\,\,\rm{Hz}, \rm{wind,}\\
 \end{array} \right.
 \label{eqn:sc:nu_a-ic<}
\end{equation}
while the expression for $\nu_{a,>}$ is
\begin{equation}
\nu_{a,>}^{\rm{IC}}=\left\{ \begin{array}{l}
1.48\times10^{18}(1+z)^{-1}\epsilon_{e,-0.5}^{-1.94}\epsilon_{B,-2.5}^{-1.10}E_{cm,53}^{-0.85}n^{-1.37}(\displaystyle\frac{t}{t_{cm}})^{-(17.9+\epsilon)/[3.1(4-\epsilon)]}\,\,\rm{Hz}, \phantom{ss}  \rm{ISM,} \\
4.99\times10^{17}(1+z)^{-1}\epsilon_{e,-0.5}^{0.11}\epsilon_{B,-2.5}^{-0.42}E_{cm,53}^{0.52}A_{\ast}^{-1.37}(\displaystyle\frac{t}{t_{cm}})^{-(9.4-3.1\epsilon)/[3.1(2-\epsilon)]}\,\,\rm{Hz},  \phantom{s}  \rm{wind,}\\
 \end{array} \right.
  \label{eqn:sc:nu_a-ic>-1}
\end{equation}
for $p=2.2$, and
\begin{equation}
\nu_{a,>}^{\rm{IC}}=\left\{ \begin{array}{l}
1.35\times10^{18}(1+z)^{-1}\epsilon_{e,-0.5}^{-1.97}\epsilon_{B,-2.5}^{-1.13}E_{cm,53}^{-0.88}n^{-1.41}(\displaystyle\frac{t}{t_{cm}})^{-(18.8+\epsilon)/[3.2(4-\epsilon)]}\,\,\rm{Hz}, \phantom{ss}  \rm{ISM,} \\
9.35\times10^{17}(1+z)^{-1}\epsilon_{e,-0.5}^{0.15}\epsilon_{B,-2.5}^{-0.42}E_{cm,53}^{0.53}A_{\ast}^{-1.41}(\displaystyle\frac{t}{t_{cm}})^{-(4.9-1.6\epsilon)/[1.6(2-\epsilon)]}\,\,\rm{Hz},  \phantom{s}  \rm{wind,}\\
 \end{array} \right.
  \label{eqn:sc:nu_a-ic>-2}
\end{equation}
for $p=2.4$. As we can see, $\nu_{a}^{\rm{IC}}$ is below the X-ray
frequency $\nu\sim10^{18}$ Hz for typical parameters. We thus do
not consider this frequency in our estimation of the IC component
in the X-ray light curve in the slow-cooling phase.

The inverse Compton spectrum is
\begin{equation}
F_{\nu}^{\rm{IC}}=\left\{
\begin{array}{l}
\displaystyle(\frac{\nu}{\nu_{m}^{\rm{IC}}})^{1/3}F_{\nu,\rm{max}}^{\rm{IC}}\propto(\frac{t}{t_{cm}})^{(12-7k)/[3(m+1)]},
\phantom{sssssssssssssssssssssssssssssssssss}  \nu<\nu_{m}^{\rm{IC}}, \\
\displaystyle(\frac{\nu}{\nu_{m}^{\rm{IC}}})^{-(p-1)/2}F_{\nu,\rm{max}}^{\rm{IC}}\propto(\frac{t}{t_{cm}})^{[16-9k+2m-p(6m+k)]/[4(m+1)]},   \phantom{sssssssssssssssss}  \nu_{m}^{\rm{IC}}<\nu<\nu_{c}^{\rm{IC}},\\
\displaystyle(\frac{\nu}{\nu_{c}^{\rm{IC}}})^{-p/2}(\frac{\nu_{c}^{\rm{IC}}}{\nu_{m}^{\rm{IC}}})^{-(p-1)/2}F_{\nu,\rm{max}}^{\rm{IC}}\propto(\frac{t}{t_{cm}})^{[6(2-k)(4-p)+p(6mp+pk-20m-4)]/[4(4-p)(m+1)]},   \phantom{}\,  \nu_{c}^{\rm{IC}}<\nu,\\
 \end{array} \right.
\label{eqn:sc:SSC-spectrum}
\end{equation}
where we have neglected the logarithmic term for
$\nu>\nu_{c}^{\rm{IC}}$ and also do not consider the lowest
spectral segment below $\nu_{a}^{\rm{IC}}$ for simplicity. The
relation between the peak flux density of the SSC spectral
component and that of the synchrotron component is (Sari \& Esin
2001)
\begin{equation}
F_{\nu,\rm{max}}^{\rm{IC}}\approx4x_{0}\frac{(p-1)(p+1/3)}{(p-1/3)(p+1)^2}(\sigma_{\rm{T}}Rn)F_{\nu,\rm{max}}.
\end{equation}
The critical frequency corresponding to the crossing point of the
synchrotron spectral component and the SSC component is
\begin{equation}
\nu_{\times}^{\rm{IC}}=\left\{
\begin{array}{l}
\nu_{\times,<}^{\rm{IC}}\equiv\nu_{m}^{\rm{IC}}[c_2\displaystyle\frac{\epsilon_{B}}{\epsilon_{e}}(\frac{\gamma_{c}}{\gamma_{e,\rm{min}}})^{4}(2\gamma_{c}\gamma_{e,\rm{min}})^{2-p}]^{3/(2+3p)}, \phantom{s} \rm{if} \phantom{ss} \nu_{\times}^{\rm{IC}}<\nu_{m}^{\rm{IC}}, \\
\nu_{\times,>}^{\rm{IC}}\equiv\nu_{m}^{\rm{IC}}[c_2\displaystyle\frac{\epsilon_{B}}{\epsilon_{e}}(\frac{\gamma_{c}}{\gamma_{e,\rm{min}}})^{4}(2\gamma_{c}\gamma_{e,\rm{min}})^{2-p}], \phantom{ssssssss} \rm{if} \phantom{ss} \nu_{m}^{\rm{IC}}<\nu_{\times}^{\rm{IC}}<\nu_{c}^{\rm{IC}}, \\
 \end{array} \right.
\label{eqn:sc:nu-ic}
\end{equation}
where we include the coefficient
$c_2=\displaystyle\frac{(1-\epsilon)^2(p-1/3)^2(p+1)^4}{9x_{0}^2(4-k-\epsilon)^2(p-2)^2(p+1/3)^2}$,
which is much larger than unity (by at least one order of
magnitude), but was neglected in equation (5.1) of Sari \& Esin
(2001). Since $\displaystyle\frac{3}{2+3p}$ is always smaller than
unity, $\nu_{\times}^{\rm{IC}}$ can be determined directly by
$\nu_{\times}^{\rm{IC}}=\rm{max}\{\nu_{\times,<}^{\rm{IC}},\nu_{\times,>}^{\rm{IC}}\}$
without judging whether $\nu_{\times}^{\rm{IC}}<\nu_{m}^{\rm{IC}}$
or not.

We have numerically calculated the temporal evolution of
$\nu_{\times,<}^{\rm{IC}}$ and $\nu_{\times,>}^{\rm{IC}}$. {\em In
the ISM case}, the expression for $\nu_{\times,<}^{\rm{IC}}$ is
\begin{equation}
\nu_{\times,<}^{\rm{IC}}=\left\{
\begin{array}{l}
3.5\times 10^{19}(1+z)^{-1}\epsilon_{e,-0.5}^{-3.08}\epsilon_{B,-2.5}^{-1.35}E_{cm,53}^{-1.47}n^{-2.43}(\displaystyle\frac{t}{t_{cm}})^{(190\epsilon-754)/[129(4-\epsilon)]}\,{\rm{Hz}}, p=2.2, \\
7.3\times 10^{18}(1+z)^{-1}\epsilon_{e,-0.5}^{-3.04}\epsilon_{B,-2.5}^{-1.33}E_{cm,53}^{-1.43}n^{-2.37}(\displaystyle\frac{t}{t_{cm}})^{(135\epsilon-522)/[92(4-\epsilon)]}\,{\rm{Hz}}, \phantom{s} p=2.4, \\
 \end{array} \right.
 \label{eqn:sc:ism:nu<-ic}
\end{equation}
while the expression for $\nu_{\times,>}^{\rm{IC}}$ is
\begin{equation}
\nu_{\times,>}^{\rm{IC}}=\left\{
\begin{array}{l}
1.7\times 10^{18}(1+z)^{-1}\epsilon_{e,-0.5}^{-3.7}\epsilon_{B,-2.5}^{-0.6}E_{cm,53}^{-1.4}n^{-2.3}(\displaystyle\frac{t}{t_{cm}})^{(0.4+38\epsilon)/[9(4-\epsilon)]}\,{\rm{Hz}},\phantom{s} p=2.2, \\
1.8\times 10^{16}(1+z)^{-1}\epsilon_{e,-0.5}^{-3.65}\epsilon_{B,-2.5}^{-0.45}E_{cm,53}^{-1.3}n^{-2.1}(\displaystyle\frac{t}{t_{cm}})^{(1.2+4.5\epsilon)/(4-\epsilon)}\,{\rm{Hz}},\phantom{ss} p=2.4. \\
 \end{array} \right.
  \label{eqn:sc:ism:nu>-ic}
\end{equation}
We can see that $\nu_{\times}^{\rm{IC}}$ decreases first with
$\nu_{\times}^{\rm{IC}}=\nu_{\times,<}^{\rm{IC}}$, then increases
with $\nu_{\times}^{\rm{IC}}=\nu_{\times,>}^{\rm{IC}}$. The time
when $\nu_{\times}^{\rm{IC}}$ reaches its minimum,
$\nu_{\times,<}^{\rm{IC}}=\nu_{\times,>}^{\rm{IC}}=\nu_{m}^{\rm{IC}}$,
is
\begin{equation}
t_{\times,m}^{\rm{IC}}=\left\{
\begin{array}{l}
4.3\epsilon_{e,-1}^{0.39}\epsilon_{B,-2.5}^{-0.47}E_{cm,53}^{-0.04}n^{-0.08}t_{cm},\phantom{sss} \rm{if}\,\,\epsilon=0.1, \\
5.1\epsilon_{e,-0.5}^{0.34}\epsilon_{B,-2.5}^{-0.41}E_{cm,53}^{-0.04}n^{-0.07}t_{cm},\phantom{ss} \rm{if}\,\,\epsilon=0.32, \\
 \end{array} \right.
\end{equation}
for $p=2.2$, and
\begin{equation}
t_{\times,m}^{\rm{IC}}=\left\{
\begin{array}{l}
17.6\epsilon_{e,-1}^{0.33}\epsilon_{B,-2.5}^{-0.48}E_{cm,53}^{-0.07}n^{-0.15}t_{cm},\phantom{sss} \rm{if}\,\,\epsilon=0.1, \\
16.7\epsilon_{e,-0.5}^{0.29}\epsilon_{B,-2.5}^{-0.41}E_{cm,53}^{-0.06}n^{-0.13}t_{cm},\phantom{ss} \rm{if}\,\,\epsilon=0.32, \\
 \end{array} \right.
\end{equation}
for $p=2.4$. The IC component could appear in the X-ray afterglow
only if the minimum of $\nu_{\times}^{\rm{IC}}$ is less
than $\nu=10^{18}\nu_{18}$ Hz, which leads to a lower limit on
the ambient density $n$. We obtain the lower limit of $n$ as
\begin{equation}
n\gtrsim\left\{
\begin{array}{l}
8.6(1+z)^{-0.43}\nu_{18}^{-0.43}\epsilon_{e,-1}^{-1.58}\epsilon_{B,-2.5}^{-0.29}E_{cm,53}^{-0.61}\,\rm{cm}^{-3},\phantom{ss} \rm{if}\,\,\epsilon=0.1, \\
1.6(1+z)^{-0.43}\nu_{18}^{-0.43}\epsilon_{e,-0.5}^{-1.54}\epsilon_{B,-2.5}^{-0.32}E_{cm,53}^{-0.61}\,\rm{cm}^{-3},\phantom{ss} \rm{if}\,\,\epsilon=0.32, \\
 \end{array} \right.
\end{equation}
for $p=2.2$, and
\begin{equation}
n\gtrsim\left\{
\begin{array}{l}
1.9(1+z)^{-0.46}\nu_{18}^{-0.46}\epsilon_{e,-1}^{-1.63}\epsilon_{B,-2.5}^{-0.30}E_{cm,53}^{-0.62}\,\rm{cm}^{-3},\phantom{sss} \rm{if}\,\,\epsilon=0.1, \\
0.5(1+z)^{-0.46}\nu_{18}^{-0.46}\epsilon_{e,-0.5}^{-1.57}\epsilon_{B,-2.5}^{-0.33}E_{cm,53}^{-0.61}\,\rm{cm}^{-3},\phantom{sss} \rm{if}\,\,\epsilon=0.32, \\
 \end{array} \right.
\end{equation}
for $p=2.4$. This lower limit of $n$ for the emergence of IC
component in the X-ray afterglow in the slow cooling phase is
typically in the range of $1-10$ cm$^{-3}$ (Sari \& Esin 2001;
Panaitescu \& Kumar 2000; Zhang \& M\'{e}sz\'{a}ros 2001).
However, the true lower limit of $n$ is even smaller than that
given in the above equations, since we have neglected the case of
$\nu_{\times}^{\rm{IC}}>\nu_{c}^{\rm{IC}}$. The spectral segment
when $\nu_{\times}^{\rm{IC}}>\nu_{c}^{\rm{IC}}$ is oversimplified
by a single power law approximation. In fact, the logarithmic term
dominates at higher frequencies. The true evolution of
$\nu_{\times}^{\rm{IC}}$ is always decreasing, although the
decreasing rate is slowed at late times (Sari \& Esin 2001).

{\em In the stellar wind case}, the expression for
$\nu_{\times,<}^{\rm{IC}}$ is
\begin{equation}
\nu_{\times,<}^{\rm{IC}}=\left\{
\begin{array}{l}
4.4\times 10^{18}(1+z)^{-1}\epsilon_{e,-0.5}^{0.56}\epsilon_{B,-2.5}^{-0.13}E_{cm,53}^{0.97}A_{\ast}^{-2.43}(\displaystyle\frac{t}{t_{cm}})^{-(127+61\epsilon)/[129(2-\epsilon)]}\,{\rm{Hz}},\phantom{s} p=2.2, \\
3.5\times 10^{18}(1+z)^{-1}\epsilon_{e,-0.5}^{0.51}\epsilon_{B,-2.5}^{-0.14}E_{cm,53}^{0.93}A_{\ast}^{-2.37}(\displaystyle\frac{t}{t_{cm}})^{-[43(2+\epsilon)]/[92(2-\epsilon)]}\,{\rm{Hz}},\phantom{sss} p=2.4, \\
 \end{array} \right.
  \label{eqn:sc:wind:nu<-ic}
\end{equation}
while the expression for $\nu_{\times,>}^{\rm{IC}}$ is
\begin{equation}
\nu_{\times,>}^{\rm{IC}}=\left\{
\begin{array}{l}
7.2\times 10^{17}(1+z)^{-1}\epsilon_{e,-0.5}^{-0.25}\epsilon_{B,-2.5}^{0.55}E_{cm,53}^{0.9}A_{\ast}^{-2.3}(\displaystyle\frac{t}{t_{cm}})^{(41.8-29\epsilon)/[9(2-\epsilon)]}\,{\rm{Hz}},\phantom{ss} p=2.2, \\
3.0\times 10^{16}(1+z)^{-1}\epsilon_{e,-0.5}^{-0.5}\epsilon_{B,-2.5}^{0.6}E_{cm,53}^{0.8}A_{\ast}^{-2.1}(\displaystyle\frac{t}{t_{cm}})^{(5.4-3.5\epsilon)/(2-\epsilon)}\,{\rm{Hz}},\phantom{ssss} p=2.4. \\
 \end{array} \right.
   \label{eqn:sc:wind:nu>-ic}
\end{equation}
The time when $\nu_{\times}^{\rm{IC}}$ reaches its minimum,
$\nu_{\times,<}^{\rm{IC}}=\nu_{\times,>}^{\rm{IC}}=\nu_{m}^{\rm{IC}}$,
is
\begin{equation}
t_{\times,m}^{\rm{IC}}=\left\{
\begin{array}{l}
1.4\epsilon_{e,-1}^{0.29}\epsilon_{B,-2.5}^{-0.24}E_{cm,53}^{0.02}A_{\ast}^{-0.05}t_{cm},\phantom{sss} \rm{if}\,\,\epsilon=0.1, \\
1.9\epsilon_{e,-0.5}^{0.29}\epsilon_{B,-2.5}^{-0.24}E_{cm,53}^{0.02}A_{\ast}^{-0.05}t_{cm},\phantom{ss} \rm{if}\,\,\epsilon=0.32, \\
 \end{array} \right.
\end{equation}
for $p=2.2$, and
\begin{equation}
t_{\times,m}^{\rm{IC}}=\left\{
\begin{array}{l}
3.1\epsilon_{e,-1}^{0.32}\epsilon_{B,-2.5}^{-0.23}E_{cm,53}^{0.04}A_{\ast}^{-0.09}t_{cm},\phantom{sss} \rm{if}\,\,\epsilon=0.1, \\
4.4\epsilon_{e,-0.5}^{0.32}\epsilon_{B,-2.5}^{-0.23}E_{cm,53}^{0.04}A_{\ast}^{-0.08}t_{cm},\phantom{ss} \rm{if}\,\,\epsilon=0.32, \\
 \end{array} \right.
\end{equation}
for $p=2.4$. The emergence of the IC component in the X-ray
afterglow requires that the minimum of $\nu_{\times}^{\rm{IC}}$ is
lower than the X-ray frequency $\nu=10^{18}\nu_{18}$ Hz, which
leads to a lower limit on $A_{\ast}$, i.e.
\begin{equation}
A_{\ast}\gtrsim\left\{
\begin{array}{l}
1.32(1+z)^{-0.42}\nu_{18}^{-0.42}\epsilon_{e,-1}^{0.17}E_{cm,53}^{0.40},\phantom{sssssssss} \rm{if}\,\,\epsilon=0.1, \\
1.55(1+z)^{-0.42}\nu_{18}^{-0.42}\epsilon_{e,-0.5}^{0.15}\epsilon_{B,-2.5}^{0.01}E_{cm,53}^{0.40},\phantom{ss} \rm{if}\,\,\epsilon=0.32, \\
 \end{array} \right.
\end{equation}
for $p=2.2$, and
\begin{equation}
A_{\ast}\gtrsim\left\{
\begin{array}{l}
1.03(1+z)^{-0.43}\nu_{18}^{-0.43}\epsilon_{e,-1}^{0.15}E_{cm,53}^{0.39},\phantom{sss} \rm{if}\,\,\epsilon=0.1, \\
1.13(1+z)^{-0.43}\nu_{18}^{-0.43}\epsilon_{e,-0.5}^{0.14}E_{cm,53}^{0.39},\phantom{ss} \rm{if}\,\,\epsilon=0.32, \\
 \end{array} \right.
\end{equation}
for $p=2.4$. The above constraint on $A_{\ast}$ is insensitive to
other physical parameters. Together with the constraint on
$A_{\ast}$ for $t<t_{cm}$, we conclude that the contribution of
the IC component to the X-ray afterglow is insignificant and can
be neglected for $A_{\ast}\lesssim1$ as indicated from
observations of Wolf-Rayet stars and fittings to some GRB
afterglows (Chevalier, Li, \& Fransson 2004; Panaitescu \& Kumar
2001, 2002).

\section{Afterglow light curves of semi-radiative blast waves}
\label{sec:lightcurves} We assume below that the physical
parameters do not deviate significantly from that chosen in
previous sections. The contamination of the IC component in the
high frequency afterglow (e.g. the soft X-ray afterglow) is not considered
for simplicity. However, the IC emissions can be inferred from
equations (\ref{eqn:fc:SSC-spectrum}) and
(\ref{eqn:sc:SSC-spectrum}). Under these assumptions, the light
curve at an observing frequency $\nu$ can be determined by
comparing the frequency with the critical frequencies $\nu_{cm}$ and
$\nu_{ac}$, where $\nu_{ac}$ is the SSA/cooling frequency at
$t_{ac}$ and can be calculated from equations
(\ref{eqn:fc:spectrum}), (\ref{eqn:fc:ISM:t_ac}) and
(\ref{eqn:fc:wind:t_ac}). Roughly, there are three types of
afterglow light curves in various frequency ranges separated by
these two critical frequencies. A careful inspection of the order
of the transition time $t_{cm}$ and the crossing times $t_{a}$,
$t_{c}$, $t_{m}$ gives four types of light curves for both the
ISM case and the stellar wind case. The crossing times $t_{a}$,
$t_{c}$ and $t_{m}$ correspond to the times when the frequencies
$\nu_{a}$, $\nu_{c}$ and $\nu_{m}$ equals the observing frequency,
respectively.

\begin{figure}
\plotone{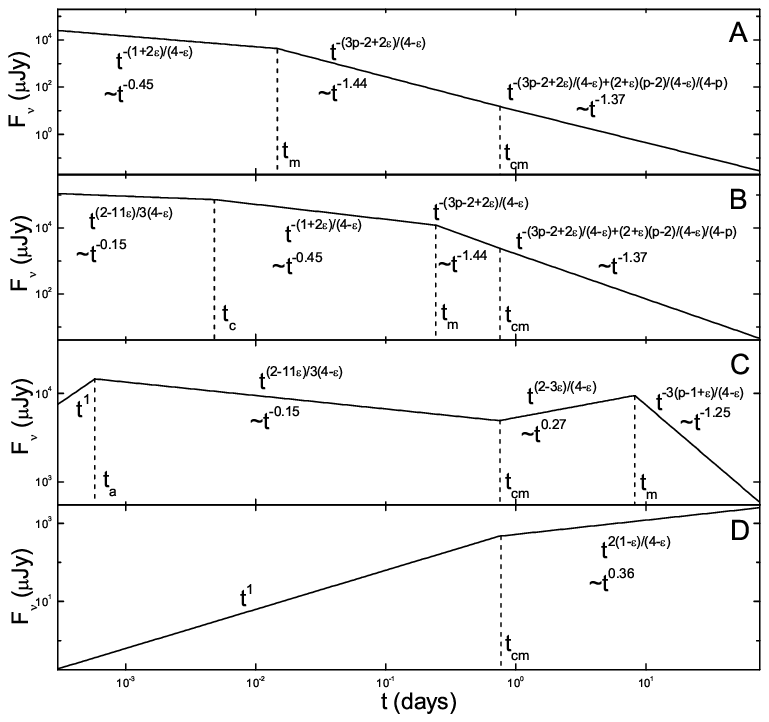} \caption{Characteristic afterglow light curves
from a semi-radiative blast wave in an interstellar medium (ISM)
in various frequency ranges. The panels are ordered from high
frequency (type A) to low frequency (type D). X-ray and optical
light curves are typically of types A and B, while sub-millimeter
and radio light curves are typically of types C and D. The
physical parameters for these light curves are $E_{cm}=10^{53}$
ergs, $n=1$ cm$^{-3}$, $\epsilon=\epsilon_{e}=1/3$,
$\epsilon_{B}=10^{-2.5}$, $p=2.2$. The event is assumed to be at
redshift of $z=1$. \label{fig:ISM:lightcurves}}
\end{figure}

{\em In the case of ISM}, the orders of these times are (A)
$t_{c}<t_{a}<t_{m}<t_{cm}$ for $\nu>\nu_{ac}$; (B)
$t_{a}<t_{c}<t_{m}<t_{cm}$ for $\nu_{cm}<\nu<\nu_{ac}$; (C)
$t_{a}<t_{cm}<t_{m}<t_{c}$ for $\nu_{a}(t_{cm})<\nu<\nu_{cm}$; (D)
$t_{cm}<t_{a}<t_{m}<t_{c}$ for $\nu<\nu_{a}(t_{cm})$. Here
$\nu_{a}(t_{cm})$ is the SSA frequency at $t_{cm}$. Using the
equations (\ref{eqn:fc:syn-spectrum-1}),
(\ref{eqn:fc:syn-spectrum-2}), and (\ref{eqn:sc:syn-spectrum}) in
the case of $k=0$, the light curves in each case can be
constructed as (A) $F_{\nu}\propto t^{1}\nu^{2}$ ($t<t_{c}$),
$F_{\nu}\propto t^{(5-2\epsilon)/(4-\epsilon)}\nu^{5/2}$ ($t_{c}$,
$t_{a}$), $F_{\nu}\propto
t^{-(1+2\epsilon)/(4-\epsilon)}\nu^{-1/2}$ ($t_{a}$, $t_{m}$),
$F_{\nu}\propto t^{-(3p-2+2\epsilon)/(4-\epsilon)}\nu^{-p/2}$
($t_{m}$, $t_{cm}$), $F_{\nu}\propto
t^{-(3p-2+2\epsilon)/(4-\epsilon)+(2+\epsilon)(p-2)/(4+\epsilon)/(4-p)}\nu^{-p/2}$
($t>t_{cm}$); (B) $F_{\nu}\propto t^{1}\nu^{2}$ ($t<t_{a}$),
$F_{\nu}\propto t^{(2-11\epsilon)/3(4-\epsilon)}\nu^{1/3}$
($t_{a}$, $t_{c}$), $F_{\nu}\propto
t^{-(1+2\epsilon)/(4-\epsilon)}\nu^{-1/2}$ ($t_{c}$, $t_{m}$),
$F_{\nu}\propto t^{-(3p-2+2\epsilon)/(4-\epsilon)}\nu^{-p/2}$
($t_{m}$, $t_{cm}$), $F_{\nu}\propto
t^{-(3p-2+2\epsilon)/(4-\epsilon)+(2+\epsilon)(p-2)/(4+\epsilon)/(4-p)}\nu^{-p/2}$
($t>t_{cm}$); (C) $F_{\nu}\propto t^{1}\nu^{2}$ ($t<t_{a}$),
$F_{\nu}\propto t^{(2-11\epsilon)/3(4-\epsilon)}\nu^{1/3}$
($t_{a}$, $t_{cm}$), $F_{\nu}\propto
t^{(2-3\epsilon)/(4-\epsilon)}\nu^{1/3}$ ($t_{cm}$, $t_{m}$),
$F_{\nu}\propto t^{-3(p-1+\epsilon)/(4-\epsilon)}\nu^{-(p-1)/2}$
($t_{m}$, $t_{c}$), $F_{\nu}\propto
t^{-(3p-2+2\epsilon)/(4-\epsilon)+(2+\epsilon)(p-2)/(4+\epsilon)/(4-p)}\nu^{-p/2}$
($t>t_{c}$); (D) $F_{\nu}\propto t^{1}\nu^{2}$ ($t<t_{cm}$),
$F_{\nu}\propto t^{2(1-\epsilon)/(4-\epsilon)}\nu^{2}$ ($t_{cm}$,
$t_{a}$), $F_{\nu}\propto t^{(2-3\epsilon)/(4-\epsilon)}\nu^{1/3}$
($t_{a}$, $t_{m}$), $F_{\nu}\propto
t^{-3(p-1+\epsilon)/(4-\epsilon)}\nu^{-(p-1)/2}$ ($t_{m}$,
$t_{c}$), $F_{\nu}\propto
t^{-(3p-2+2\epsilon)/(4-\epsilon)+(2+\epsilon)(p-2)/(4+\epsilon)/(4-p)}\nu^{-p/2}$
($t>t_{c}$). The light curves in the ISM case are illustrated in
Figure \ref{fig:ISM:lightcurves}. The crossing times $t_{c}$ and
$t_{a}$ in case A and $t_{a}$ in case B occur very early in high
observing frequencies, while $t_{c}$ in case C and $t_{a}$,
$t_{m}$ and $t_{c}$ in case D occur very late in low observing
frequencies. We thus neglect these crossing times in the figure.
As indicated in Figure \ref{fig:ISM:lightcurves}, the radiation
efficiency, $\epsilon$, has a marked effect on the afterglow light
curves. It changes the temporal decaying index $\alpha$ (defined
as $F_{\nu}\propto t^{-\alpha}$) of the light curve significantly.
For illustration, we adopt $\epsilon=\epsilon_{e}=1/3$ and $p=2.2$
in the following. The initial slowly increasing light curve
segment, $F_{\nu}\propto t^{1/6}$, predicted in the standard
adiabatic blast wave model will changes to be a slowly decreasing
one, $F_{\nu}\propto t^{(2-11\epsilon)/3(4-\epsilon)}\sim
t^{-0.15}$, as shown in the early segment of case C. This makes
the sub-millimeter afterglow less competitive to distinguish
between the ISM and the stellar wind, as proposed by Panaitescu \&
Kumar (2000), if the observations are not frequent enough. At the
optical wavelength, the early light curve behaves typically as
$F_{\nu}\propto t^{-(1+2\epsilon)/(4-\epsilon)}\sim t^{-0.45}$
rather than $F_{\nu}\propto t^{-1/4}$ in the adiabatic case. When
$\nu_{m}$ crosses the observing frequency, i.e. $t>t_{m}$, the
optical and X-ray light curves decays as $F_{\nu}\propto
t^{-(3p-2+2\epsilon)/(4-\epsilon)}\sim t^{-1.44}$, as shown in
cases B and A. It should be noted that many X-ray afterglow light
curves and a considerable fraction of optical afterglow light
curves have temporal decaying indices steeper than predicted
$(3p-2)/4\sim 1.15$ by the standard adiabatic model (for
$p\approx2.2$). The observed decaying indices in the X-ray
afterglow light curves are $\langle\alpha_{X}\rangle=1.33\pm0.38$,
while the median value of the observed X-ray spectral indices
$\beta_{X}$, $F_{\nu}\propto \nu^{-\beta_{X}}$, is $\sim1.05$
(Berger, Kulkarni, \& Frail 2003; De Pasquale et al. 2003).
Assuming the X-ray frequency is above the cooling frequency and
the spectral index is $\beta_{X}=p/2$, the measured $p$ is
consistent with the standard value of the index of electron energy
distribution, $p=2.2-2.3$, predicted in the relativistic shock
acceleration mechanism (see Achterberg et al. 2001 and reference
therein). However, the observed mean temporal decaying index
$\langle\alpha_{X}\rangle$ requires a relatively larger $\langle
p\rangle$, $\sim2.44$, provided the shock is adiabatic. There are
several caveats on the observations of the X-ray afterglows.
First, the temporal behavior of the X-ray afterglow is hardly
influenced by the equal arrival time surface effect, which will
mix the earlier light from high latitudes into the present light.
This effect is important especially for high observing
frequencies, e.g. the optical and X-ray. The profile of surface
emissivity of the relativistic shock is ring-like in these high
frequencies (Sari 1998). This will moderately slow down the
decreasing rate of the afterglow after $t_{m}$ for the theoretical
light curve A. However, the X-ray afterglow is immune to this
effect because its $t_{m}$ is very early and the observed
decreasing index $\alpha_{X}$ is based on the observations
typically several hours after the burst. Second, the measured
$\beta_{X}$ is reliable since the X-ray absorption in the medium
along the line of sight takes place at $\nu\lesssim 1$ keV while
the observing window is $\sim2-10$ keV. Lastly, one should be
cautious when interpreting the property of X-ray afterglow with
the synchrotron radiation mechanism, because the X-ray afterglow
may be contaminated by the synchrotron-self-Compton component.
However, there have so far been only a few X-ray afterglows that
were confirmed to have the IC components. Therefore, the radiative
corrections to the afterglow light curves must be taken into
account based on the observations. As can be seen in Figure
\ref{fig:ISM:lightcurves}, the light curve at high frequency (type
A and B) flattens when the afterglow enters the slow-cooling
phase, $t>t_{cm}$. At the transition time $t_{cm}$, the spectrum
nearby the observing frequency changes from $\nu_{c}<\nu_{m}<\nu$
to $\nu_{m}<\nu_{c}<\nu$, while the expressions for the flux
density are the same, i.e.
$F_{\nu}=F_{\nu,\max}\nu_{m}^{(p-1)/2}\nu_{c}^{1/2}\nu^{-p/2}$.
The flattening of the light curve results from the Compton
parameter $Y$ in the flux density, i.e. $F_{\nu}\propto Y^{-1}$,
since $\nu_{c}\propto(1+Y)^{-2}\approx Y^{-2}$. The Compton
parameter $Y$ in the slow-cooling phase decreases slowly, contrary
to its constancy in the earlier fast-cooling phase. From equation
(\ref{eqn:sc:spectrum}) for $Y$ in the slow-cooling phase and
adopting $k=0$, the change of the temporal index around $t_{cm}$
is $\Delta\alpha=(2+\epsilon)(p-2)/[(4-\epsilon)(4-p)]$, which is
shown in the last segments of panels A and B in Figure
\ref{fig:ISM:lightcurves}. Note that since Sari et al. (1998) did
not discuss IC cooling, there is no related segment in their
$\epsilon=0$ light curves.

\begin{figure}
\plotone{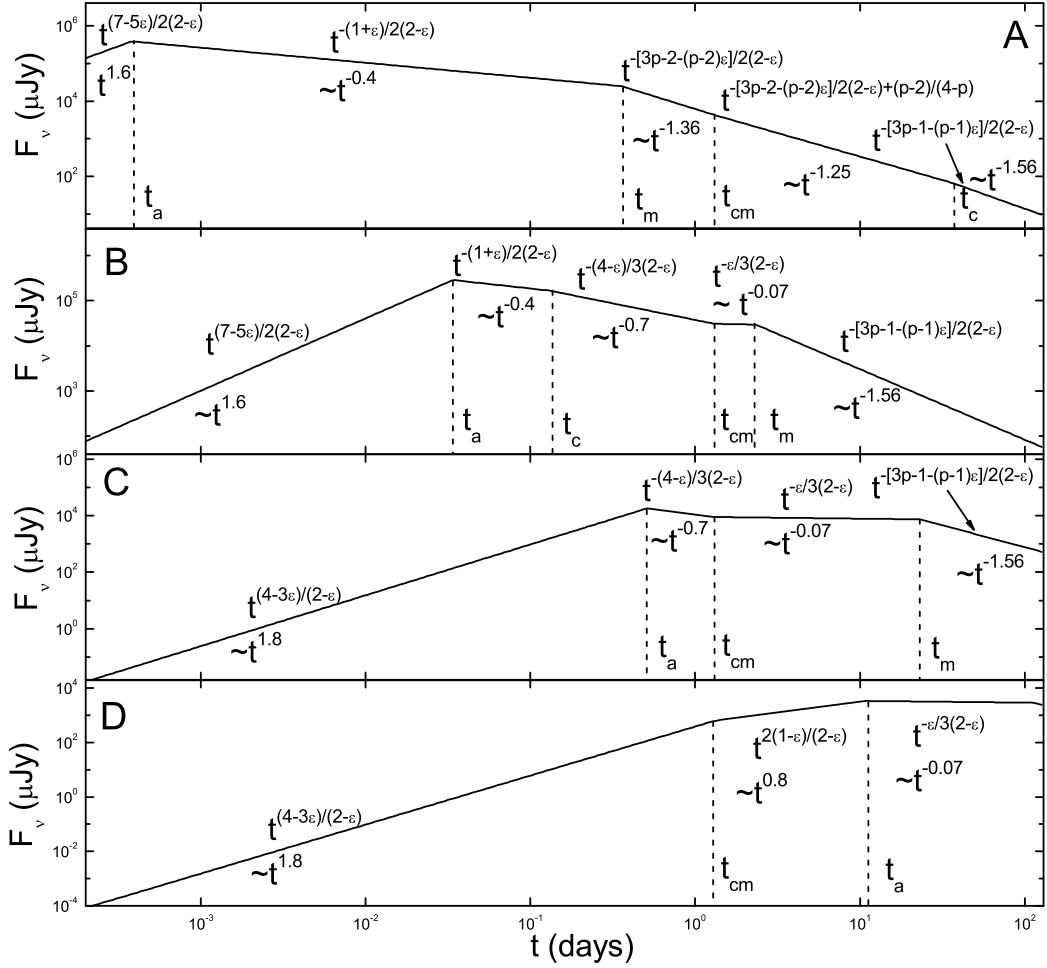} \caption{Characteristic afterglow light curves
from a semi-radiative blast wave in a stellar wind in various
frequency ranges. The panels are ordered from high frequency (type
A) to low frequency (type D). X-ray, optical, and infrared light
curves are typically of type A, sub-millimeter light curve is
typically of type B, while radio light curve is typically of types
C and D. The physical parameters for these light curves are
$E_{cm}=10^{53}$ ergs, $A_{\ast}=1$, $\epsilon=\epsilon_{e}=1/3$,
$\epsilon_{B}=10^{-2.5}$, $p=2.2$. The event is assumed to be at
redshift of $z=1$.\label{fig:wind:lightcurves}}
\end{figure}

{\em In the case of stellar wind}, the orders of the crossing
times are (A) $t_{a}<t_{m}<t_{cm}<t_{c}$ for $\nu>\nu_{cm}$; (B)
$t_{a}<t_{c}<t_{cm}<t_{m}$ for $\nu_{ac}<\nu<\nu_{cm}$; (C)
$t_{c}<t_{a}<t_{cm}<t_{m}$ for $\nu_{a}(t_{cm})<\nu<\nu_{ac}$; (D)
$t_{c}<t_{cm}<t_{a}<t_{m}$ for $\nu<\nu_{a}(t_{cm})$. Using the
equations (\ref{eqn:fc:syn-spectrum-1}),
(\ref{eqn:fc:syn-spectrum-2}), and (\ref{eqn:sc:syn-spectrum}) in
the case of $k=2$, the light curves in each case can be
constructed as (A) $F_{\nu}\propto
t^{(7-5\epsilon)/2(2-\epsilon)}\nu^{5/2}$ ($t<t_{a}$),
$F_{\nu}\propto t^{-(1+\epsilon)/2(2-\epsilon)}\nu^{-1/2}$
($t_{a}$, $t_{m}$), $F_{\nu}\propto
t^{-[3p-2-(p-2)\epsilon]/2(2-\epsilon)}\nu^{-p/2}$ ($t_{m}$,
$t_{cm}$), $F_{\nu}\propto
t^{-[3p-2-(p-2)\epsilon]/2(2-\epsilon)+(p-2)/(4-p)}\nu^{-p/2}$
($t_{cm}$, $t_{c}$), $F_{\nu}\propto
t^{-[3p-1-(p-1)\epsilon]/2(2-\epsilon)}\nu^{-(p-1)/2}$
($t>t_{c}$); (B) $F_{\nu}\propto
t^{(7-5\epsilon)/2(2-\epsilon)}\nu^{5/2}$ ($t<t_{a}$),
$F_{\nu}\propto t^{-(1+\epsilon)/2(2-\epsilon)}\nu^{-1/2}$
($t_{a}$, $t_{c}$), $F_{\nu}\propto
t^{-(4-\epsilon)/3(2-\epsilon)}\nu^{1/3}$ ($t_{c}$, $t_{cm}$),
$F_{\nu}\propto t^{-\epsilon/3(2-\epsilon)}\nu^{1/3}$ ($t_{cm}$,
$t_{m}$), $F_{\nu}\propto
t^{-[3p-1-(p-1)\epsilon]/2(2-\epsilon)}\nu^{-(p-1)/2}$
($t>t_{m}$); (C) $F_{\nu}\propto
t^{(7-5\epsilon)/2(2-\epsilon)}\nu^{5/2}$ ($t<t_{c}$),
$F_{\nu}\propto t^{(4-3\epsilon)/(2-\epsilon)}\nu^{2}$ ($t_{c}$,
$t_{a}$), $F_{\nu}\propto
t^{-(4-\epsilon)/3(2-\epsilon)}\nu^{1/3}$ ($t_{a}$, $t_{cm}$),
$F_{\nu}\propto t^{-\epsilon/3(2-\epsilon)}\nu^{1/3}$ ($t_{cm}$,
$t_{m}$), $F_{\nu}\propto
t^{-[3p-1-(p-1)\epsilon]/2(2-\epsilon)}\nu^{-(p-1)/2}$
($t>t_{m}$); (D) $F_{\nu}\propto
t^{(7-5\epsilon)/2(2-\epsilon)}\nu^{5/2}$ ($t<t_{c}$),
$F_{\nu}\propto t^{(4-3\epsilon)/(2-\epsilon)}\nu^{2}$ ($t_{c}$,
$t_{cm}$), $F_{\nu}\propto t^{2(1-\epsilon)/(2-\epsilon)}\nu^{2}$
($t_{cm}$, $t_{a}$), $F_{\nu}\propto
t^{-\epsilon/3(2-\epsilon)}\nu^{1/3}$ ($t_{a}$, $t_{m}$),
$F_{\nu}\propto
t^{-[3p-1-(p-1)\epsilon]/2(2-\epsilon)}\nu^{-(p-1)/2}$
($t>t_{m}$).  The light curves in the stellar wind case are
illustrated in Figure \ref{fig:wind:lightcurves}. The crossing
time $t_{c}$ in cases C and D occurs very early at low observing
frequency, while $t_{m}$ in case D occurs very late. We neglect
these crossing times in this figure. The radiation efficiency has
a significant effect on the light curves in the wind case. The
flux density in the optical/infrared light curve decays initially
with $t^{-(1+\epsilon)/2(2-\epsilon)}\sim t^{-0.4}$, rather than
$t^{-1/4}$ in the adiabatic case, which is shown in case A. For
the X-ray afterglow the crossing time $t_{m}$ when the typical
frequency $\nu_{m}$ crosses the observing frequency is much
earlier, the light curve behaves as
$t^{-[3p-2-(p-2)\epsilon]/2(2-\epsilon)}\sim t^{-1.36}$ during the
whole fast-cooling phase, which is more consistent with
observations than the adiabatic light curve (Berger, Kulkarni, \&
Frail 2003). The light curve at high frequency flattens when the
afterglow transits to the slow-cooling phase. By the same way as
in the ISM case, from equation (\ref{eqn:sc:spectrum}) for $Y$ and
adopting $k=2$, the change of the temporal index around $t_{cm}$
is $\Delta\alpha=(p-2)/(4-p)$ in the wind case. Since Chevalier \&
Li (2000) did not include IC cooling, there is no relevant
flattening segment in their $\epsilon=0$ light curves. Although
the flattening of the optical/X-ray light curve around $t_{cm}$
predicted in the inverse Compton dominated cooling regime in the
stellar wind case is more obvious than in the ISM case, the change
of the temporal decaying index is only $\Delta\alpha\sim 0.1$ for
the former case. The detailed theoretical optical light curves
taking into account the equal arrival time surface effect and the
large error bars in X-ray afterglow observations prevent us from
the identifications of such flattening.

\section{Conclusions and Discussion}
\label{sec:conclusion} In this paper, we have investigated
analytically the GRB afterglow hydrodynamics and constructed the
semi-radiative light curves realistically. We focus on the case
that the electron cooling is in the inverse Compton dominated
regime, i.e. $\epsilon_{e}\gg\epsilon_{B}$ or $Y\gg1$. The
realistic hydrodynamics is applicable for spherical blast waves
with the assumption that the electron energy equipartition factor
$\epsilon_{e}$ is not much larger than $1/3$, which seems to be
reasonable based on theoretical expectations and observations as
well. In fact, the analytical solution for afterglow hydrodynamics
is almost tenable throughout the relativistic stage when
$\epsilon_{e}\lesssim 2/3$ (see equation \ref{eqn:hydro}). The
only uncertainty is the actual evolution of the radiation
efficiency in the late slow-cooling phase. Given $p\sim2.2-2.3$,
we conclude that a constant radiation efficiency is a good
approximation for a fairly long time in the slow-cooling phase.
The transition from fast-cooling to slow-cooling happens much
later in the IC dominated cooling regime than in the purely
synchrotron cooling regime. Since the actual radiation efficiency
decreases very slowly in the slow-cooling phase, the
semi-radiative epoch is further prolonged typically by two orders
of magnitude in time, other than the whole fast-cooling phase as
commonly used. As the GRB ejecta sweeps up more and more external
medium, the Lorentz factor $\gamma$ of the shock decreases. When
$\gamma$ equals the inverse of the initial half-opening angle
$\theta_{0}$ of the GRB conical ejecta, or jet, the following
afterglow light curves deviate from the previously spherical-like
ones, and the light curves in this paper are not suitable at this
late stage. Generally, our semi-radiative afterglow light curves
hold true when the shock is relativistic after the initial
deceleration time and before the jet-like stage, with the
radiation efficiency satisfying
$\epsilon\approx\epsilon_{e}\lesssim$max$\{\displaystyle\frac{2}{2+\theta_{0}},\frac{2}{3}\}$.

The adiabatic afterglow light curves in the synchrotron dominated
cooling regime have been well studied in previous works (Sari et
al. 1998; Chevalier \& Li 2000; Granot \& Sari 2002). Sari et al.
(1998) have considered the light curves of a fully radiative
($\epsilon=1$) blast wave. However, our analytical results can not
be directly applied to this case. Actually, in the fully radiative
case, equation (\ref{eqn:hydro}) can also be directly integrated
and the hydrodynamics and the resulting light curves are the same
as derived by Sari et al. (1998). The radiative corrections to the
afterglow light curves in the wind model have been discussed by
Chevalier \& Li (2000). They had adopted the scaling laws of a
semi-radiative shock given by B\"{o}ttcher \& Dermer (2000). The
temporal exponents of the hydrodynamics and light curves in
B\"{o}ttcher \& Dermer (2000) differ from ours. In fact, the
radiation corrections for these exponents in their work are
smaller than that in our
work\footnotemark\footnotetext{\label{foot:lightcurve}The
coefficient of the $\epsilon$ term in these temporal exponents in
their work is smaller than ours by an exact factor of 2.}. Cohen,
Piran \& Sari (1998) had studied the hydrodynamics of a
semi-radiative relativistic blast wave considering a more
complicated post-shock material distribution (similar to Blandford
\& McKee 1976) than that of the simple thin-shell approximation as
in our work. However, the hydrodynamic evolution differs in all
these three treatments of Cohen et al. (1998), B\"{o}ttcher \&
Dermer (2000) and this paper. The hydrodynamic self-similarity
index $m$ we obtain lies between the other two's. Despite the
difference between the hydrodynamics due to different
approximations/treatments, we can see that the radiative
corrections to afterglow light curves are significant. For
example, the temporal decaying index of X-ray afterglows at around
$10$ hours since the main bursts varies in the range of
$1.23-1.69$ in the ISM case, and $1.21-1.58$ in the stellar wind
case for $p=2.2-2.3$ and $\epsilon_{e}=0.1-0.5$, provided that the
IC component is neglected. This range is consistent with the
observations (Berger, Kulkarni \& Frail 2003). In observations,
the value of $\epsilon_{e}$ and $p$ can be inferred for a
particular X-ray afterglow from the so called ``closure relation"
between the temporal index $\alpha_{X}$ and spectral index
$\beta_{X}$, provided that the IC component can be neglected. Such
a closure relation can be obtained from case A described in \S
\ref{sec:lightcurves} in either ISM or wind cases. The application
of this method to optical afterglows should be cautious. The early
optical light curve has a broken power law profile around $t_{m}$.
In contrary to the early X-ray afterglow whose $t_{m}$ is much
earlier and which can be regarded as a simple power law one, the
optical light curve might be affected moderately when the equal
arrival time surface effect is included. Another reason is that
the reddening in optical spectra is unknown. While the Galactic
extinction can be empirically decoupled, the extinction and
reddening within the circum-burst environment and host galaxy are
less constrained. These two facts prevent the credible
measurements of the temporal and spectral indices of optical
afterglows, respectively.

We have got the criteria for the emergence of IC components in
soft X-ray afterglows, through giving the lower limits of external
medium densities. In the ISM case, the lower limit of $n$ is
$\sim10$ cm$^{-3}$ in the fast-cooling phase, while it is $\sim 1$
cm$^{-3}$ in the slow-cooling phase (Panaitescu \& Kumar 2000;
Sari \& Esin 2001; Zhang \& M\'{e}sz\'{a}ros 2001). These are
typical densities of interstellar media in our galaxy. In the wind
case, the lower limit of the wind parameter is always
$A_{\ast}\sim 1$ (Panaitescu \& Kumar 2000). Such a critical
$A_{\ast}$ is also typical for Wolf-Rayet stars in our galaxy. It
should be noted that the wind parameter obtained from fitting
afterglows within the wind interaction model seems to be quite
small (Chevalier \& Li 1999, 2000; Panaitescu \& Kumar 2001; Dai
\& Wu 2003; Chevalier, Li, \& Fransson 2004). The contradiction
may be due to the limitation of our knowledge about the mass
losses of massive stars at the last stage before their collapses.
Taking the inferred low $A_{\ast}$ from afterglows, we draw a
tentative conclusion that the IC components in X-ray afterglows
are insignificant in the stellar wind case. For such a low
$A_{\ast}$, a wind bubble would be produced and surrounded by
either an outer giant molecular cloud, a slow wind at previous
evolutionary stage, or an extremely high pressure in a star-burst
environment (Dai \& Wu 2003; Chevalier, Li, \& Fransson 2004). The
termination shock radius of the wind bubble will be reached by the
GRB shock within hours in observer's frame. The environment of the
shock will change from wind type to uniform medium type after this
time. We have neglected such a complicated case in this work.
Recently, Yost et al. (2003) had relaxed the assumption of a
constant magnetic equipartition factor $\epsilon_{B}$ and
broadened the circum-burst medium types. They had made a detailed
comparison between the results of different assumptions as well as
different circum-burst media, and found the degeneracy of
different assumed evolutions of $\epsilon_{B}$ and different
medium types. Anyway, we adopt the constant $\epsilon_{B}$
assumption and consider the most possible medium types, i.e. the
ISM and the stellar wind.

The radiative corrections in modelling afterglows are important.
Such an effect should be taken into account seriously in analysis
of afterglows, especially when a large number of afterglows will
be observed in the \emph{Swift} era. It will affect directly the
energetics of GRB remnants and therefore the actual efficiency of
prompt GRBs (Wu et al. in preparation, which also includes the ICS
effect).

We thank the referee for his/her valuable suggestions and
comments. This work was supported by the National Natural Science
Foundation of China (grants 10233010, 10221001, 10003001, and
10473023), the Ministry of Science and Technology of China (NKBRSF
G19990754), the Special Funds for Major State Basic Research
Projects, and the Foundation for the Author of National Excellent
Doctoral Dissertation of P. R. China (Project No. 200125).

\clearpage

\begin{thebibliography}{}
\bibitem[{Achterberg} {\etal}(2001)]{agk+01}{Achterberg}, A., {Gallant}, Y. A., Kirk, J. G., \& {Guthmann}, A. W. 2001, \mnras, 328, 393
\bibitem[{Berger}, {Kulkarni} \& {Frail}(2003)]{bkf+03}{Berger}, E., {Kulkarni}, S. R., \& {Frail}, D. A. 2003, \apj, 590, 379
\bibitem[{Bj\"{o}rnsson}(2001)]{bj+01}{Bj\"{o}rnsson}, C. I. 2001, \apj, 554, 593
\bibitem[{Blandford} \& {McKee}(1976)]{bm+76}{Blandford}, R. D., \& {McKee}, C. F. 1976, Phys. Fluids, 19, 1130
\bibitem[{B\"{o}ttcher} \& {Dermer}(2000)]{bd+00}{B\"{o}ttcher}, M., \& {Dermer}, C. D. 2000, \apjl, 532, 281
\bibitem[{Cheng} \& {Lu}(2001)]{cl+01}{Cheng}, K. S., \& {Lu}, T. 2001, ChJAA, 1, 1
\bibitem[{Chevalier} \& {Li}(1999)]{cl+99}{Chevalier}, R. A., \& {Li}, Z. Y. 1999, \apjl, 520, L29
\bibitem[{Chevalier} \& {Li}(2000)]{cl+00}{Chevalier}, R. A., \& {Li}, Z. Y. 2000, \apj, 536, 195
\bibitem[{Chevalier}, {Li} \& {Fransson}(2004)]{clf+04}{Chevalier}, R. A., {Li}, Z. Y., \& {Fransson}, C. 2004, \apj, 606, 369
\bibitem[{Cohen} {\etal}(1998)]{cps+98}{Cohen}, E., {Piran}, T., \& {Sari}, R. 1998, \apj, 509, 717
\bibitem[{Dai} \& {Lu}(1998)]{dl+98}{Dai}, Z. G., \& {Lu}, T. 1998, \mnras, 298, 87
\bibitem[{Dai} \& {Lu}(1999)]{dl+99}{Dai}, Z. G., \& {Lu}, T. 1999, \apjl, 519, L155
\bibitem[{Dai} \& {Lu}(2000)]{dl+00}{Dai}, Z. G., \& {Lu}, T. 2000, \apj, 537, 803
\bibitem[{Dai} \& {Wu}(2003)]{dw+03}{Dai}, Z. G., \& {Wu}, X. F. 2003, \apjl, 591, L21
\bibitem[{Dermer} {B\"{o}ttcher} \& {Chiang}(2000)]{dbc+00}{Dermer}, C. D., {B\"{o}ttcher}, M., \& {Chiang}, J. 2000, \apj, 537, 255
\bibitem[{De Pasquale} {\etal}(2003)]{dpp+03}{De Pasquale}, M., {\etal} 2003, \apj, 592, 1018
\bibitem[{Feng} {\etal}(2002)]{fhdl+02}{Feng}, J. B., Huang, Y. F., {Dai}, Z. G., \& {Lu}, T. 2002, ChJAA, 2, 525
\bibitem[{Galama} {\etal}(1998)]{gvp+98}{Galama}, T. J., {\etal} 1998, \nat, 395, 670
\bibitem[{Galama} {\etal}(1998)]{gwb+98}{Galama}, T. J., {Wijers}, R. A. M. J., {Bremer}, M., {Groot}, P. J., {Strom}, R. G.,
                  {Kouveliotou}, C., \& {van Paradijs}, J. 1998, \apjl, 500, L97
\bibitem[{Granot} \& {Sari}(2002)]{gs+02}{Granot}, J., \& {Sari}, R. 2002, \apj, 568, 820
\bibitem[{Hjorth} {\etal}(2003)]{jsm+03}{Hjorth}, J., {\etal} 2003, \nat, 423, 847
\bibitem[{Huang} {\etal}(1999)]{hdl+99}{Huang}, Y. F., {Dai}, Z. G., \& {Lu}, T. 1999, \mnras, 309, 513
\bibitem[{Huang} {\etal}(2000)]{hgdl+00}{Huang}, Y. F., {Gou}, L. J., {Dai}, Z. G., \& {Lu}, T. 2000, \apj, 543, 90
\bibitem[{Kulkarni} {\etal}(1998)]{kfw+98}{Kulkarni}, S. R., {\etal} 1998, \nat, 395, 663
\bibitem[{Li} {\etal}(2002)]{ldl+02}{Li}, Z., {Dai}, Z. G., \& {Lu}, T. 2002, \mnras, 330, 955
\bibitem[{Lloyd-Ronning} \& {Zhang}(2004)]{lz+04}{Lloyd-Ronning}, N. M., \& {Zhang}, B. 2004, \apj, 613, 477
\bibitem[{\meszaros}, \& {Rees}(1997)]{mr+97}{\meszaros}, P., \& {Rees}, M. 1997, \apj, 476, 232
\bibitem[{\meszaros}, {Rees} \& {Wijers}(1998)]{mrw+98}{\meszaros}, P., {Rees}, M., \& {Wijers}, R. A. M. J. 1998, \apj, 499, 301
\bibitem[{\meszaros} (2002)]{mes+02}{\meszaros}, P. 2002, ARA\&A, 40, 137
\bibitem[{Paczy\'{n}ski}(1998)]{pac+92}{Paczy\'{n}ski}, B. 1998, \apjl, 494, L45
\bibitem[{Panaitescu} \& {Kumar}(2000)]{pk+00}{Panaitescu}, A., \& {Kumar}, P. 2000, \apj, 543, 66
\bibitem[{Panaitescu} \& {Kumar}(2001)]{pk+01}{Panaitescu}, A., \& {Kumar}, P. 2001, \apjl, 560, L49
\bibitem[{Panaitescu} \& {Kumar}(2002)]{pk+02}{Panaitescu}, A., \& {Kumar}, P. 2002, \apjl, 571, 779
\bibitem[{Piran}(2004)]{piran+04}{Piran}, T. 2004, Rev. Mod. Phys., in press (astro-ph/0405503)
\bibitem[{Sari}, {Piran} \& {Narayan}(1998)]{spn+98}{Sari}, R., {Piran}, T., \& {Narayan}, R. 1998, \apjl, 497, L17
\bibitem[{Sari}(1998)]{sari+98}{Sari}, R. 1998, \apjl, 494, L49
\bibitem[{Sari} \& {Esin}(2001)]{se+01}{Sari}, R., \& {Esin}, A. A. 2001, \apj, 548, 787
\bibitem[{Stanek} {\etal}(2003)]{smg+03}{Stanek}, K. Z., {\etal} 2003, \apjl, 591, L17
\bibitem[{Van Paradijs} {\etal}(2000)]{vanp+00}{Van Paradijs}, J., {Kouveliotou}, C., \& {Wijers}, R. A. M. J. 2000, ARA\&A, 38, 379
\bibitem[{Vietri}(1997)]{vie+97}{Vietri}, M. 1997, \apjl, 478, L9
\bibitem[{Wang} {\etal}(2001)]{wdl+01}{Wang}, X. Y., {Dai}, Z. G., \& {Lu}, T. 2001, \apj, 556, 1010
\bibitem[{Waxman}(1997)]{wax+97}{Waxman}, E. 1997, \apjl, 485, L5
\bibitem[{Wijers}, {Rees} \& {\meszaros}(1997)]{wrm+97}{Wijers}, R. A. M. J., {Rees}, M. J, \& {\meszaros}, P. 1997, \mnras, 288, L51
\bibitem[{Wei} \& {Lu}(1998)]{wl+98}{Wei}, D. M., \& {Lu}, T. 1998, \apj, 505, 252
\bibitem[{Wei} \& {Lu}(2000)]{wl+00}{Wei}, D. M., \& {Lu}, T. 2000, \aap, 360, L13
\bibitem[{Wooley}(1993)]{woo+93}{Woosley}, S. E. 1993, \apj, 405, 273
\bibitem[{Wu} {\etal}(2003)]{wdhl+03}{Wu}, X. F., {Dai}, Z. G., Huang, Y. F., \& {Lu}, T. 2003, \mnras, 342, 1131
\bibitem[{Yost} {\etal}(2003)]{yhsf+03}{Yost}, S. A., {Harrison}, F. A., Sari, R., \& {Frail}, D. A. 2003, \apj, 597, 459
\bibitem[{Zhang} \& {\meszaros}(2001)]{zm+01}{Zhang}, B., \& {\meszaros}, P. 2001, \apj, 559, 110
\bibitem[{Zhang} \& {\meszaros}(2004)]{zm+04}{Zhang}, B., \& {\meszaros}, P. 2004, Int. J. Mod. Phys. A, 19, 2385
\end{thebibliography}
\end{document}